\documentclass[final,1p,times]{elsarticle}

\usepackage{graphicx}
\usepackage{amssymb}

\journal{Physica D}

\begin{document}

\begin{frontmatter}


\title{The speed of Arnold diffusion}
\author[ceft]{C. Efthymiopoulos}
\ead{cefthim@academyofathens.gr}
\author[mhar]{M. Harsoula}
\ead{mharsoul@academyofathens.gr}
\address[ceft,mhar]{Research Center for Astronomy
and Applied Mathematics, Academy of Athens}

\begin{abstract}
A detailed  numerical study is presented of the slow diffusion
(Arnold diffusion) taking place around resonance crossings in nearly
integrable Hamiltonian systems of three degrees of freedom in the
so-called `Nekhoroshev regime'. The aim is to construct estimates
regarding the speed of diffusion  based on the numerical values of a
truncated form of the so-called remainder of a normalized
Hamiltonian function, and to compare them with the outcomes of
direct numerical experiments using ensembles of orbits. In this
comparison we examine, one by one, the main steps of the so-called
analytic and geometric parts of the Nekhoroshev theorem. Thus: i) we
review and implement an algorithm \cite{eft2008} for Hamiltonian
normalization in multiply resonant domains which is implemented as a
computer program making calculations up to a high normalization
order. ii) We compute the dependence of the optimal normalization
order on the small parameter $\epsilon$ in a specific model and
compare the result with theoretical estimates on this dependence.
iii) We examine in detail the consequences of assuming simple
convexity conditions for the unperturbed Hamiltonian on the geometry
of the resonances and on the phase space structure around resonance
crossings. iv) We discuss the dynamical mechanisms by which the
remainder of the optimal Hamiltonian normal form drives the
diffusion process. Through these steps, we are led to two main
results: i) We construct in our concrete example a convenient set of
variables, proposed first by Benettin and Gallavotti
\cite{bengal1986}, in which the phenomenon of Arnold diffusion in
doubly resonant domains can be clearly visualized. ii) We determine,
by numerical fitting of our data the dependence of the local
diffusion coefficient $D$ on the size  $||R_{opt}||$ of the optimal
remainder function, and we compare this with a heuristic argument
based on the assumption of normal diffusion. We find a power law
$D\propto ||R_{opt}||^{2(1+b)}$, where the constant $b$ has a small
positive value depending also on the multiplicity of the resonance
considered.
\end{abstract}

\begin{keyword}
Hamiltonian systems; Arnold diffusion;
normal forms; Nekhoroshev theorem.
\end{keyword}

\end{frontmatter}

\section{Introduction}
The study of {\it diffusion} in nearly-integrable Hamiltonian
dynamical systems of the form
\begin{equation}\label{hamgen}
H(I,\phi)=H_0(I)
+\epsilon H_1(I,\phi)
\end{equation}
where $(I,\phi)$ are n-dimensional action - angle variables and $\epsilon$ is
a small parameter, constitutes a central problem in Hamiltonian dynamical systems
theory, in view, in particular, of its multiple applications in physics and
astronomy (see \cite{lichlie1992}\cite{con2002} for an introduction, the
basic review paper \cite{chi1979}, or \cite{mor2002}\cite{cin2002}
\cite{legetal2007} for recent advanced reviews emphasizing various
aspects of this subject). It is a well established result that, if
$n>2$, and $H$ satisfies appropriate convexity and analyticity conditions
(see section 2 below), two distinct regimes characterize the laws of
diffusion as a function of $\epsilon$: for $\epsilon<\epsilon_*$, where
$\epsilon_*$ is a threshold value, the onset of the so-called `Nekhoroshev
regime' takes place \cite{nek1977}\cite{benetal1985}\cite{loch1992}
\cite{posh1993}\cite{morguz1997}\cite{mor2002}. In this case, the
Nekhoroshev theorem provides an
$O[\exp(-(\epsilon_*/\epsilon)^c)]$ upper bound for the speed of diffusion.
The exponent $c$ depends on the number of degrees of freedom $n$, while its
precise value in local domains of the action space depends also on the
multiplicity of the resonance conditions holding in such domains (see e.g.
\cite{bengal1986}\cite{posh1993}\cite{morguz1997}). Furthermore, the mechanism
of diffusion
caused by {\it transition chains}, as demonstrated in one special example
by Arnold \cite{arn1964} (see also \cite{simval2001}), is conjectured to
hold in more general systems of the form (\ref{hamgen}) (e.g. \cite{math2004};
note, however, that no formal proof of this fact has been given to date).
On the other hand, for $\epsilon>\epsilon_*$, the diffusion is driven mainly
by the mechanism of {\it resonance overlap} \cite{rosetal1966}\cite{con1967}
\cite{chi1979}. In this case, one expects a power-law dependence of the speed
of diffusion on $\epsilon$ (see e.g. \cite{chi1979}; a power law is also
found in the case of the so-called `Fast Arnold diffusion' \cite{chivech1985}).

The diffusion in weakly chaotic systems has been a subject also of
extensive numerical studies over the last three decades (some
indicative references are
\cite{kankon1989}\cite{woodetal1990}\cite{las1993}\cite{dumlas1993}
\cite{skoetal1997}\cite{eftetal1997}\cite{giocin2004}\cite{cachetal2010}).
A detailed study, however, of the very slow diffusion characterizing
the `Nekhoroshev regime' has become possible only in recent years.
In this respect, we note in particular the series of instructive
works \cite{froetal2000}
\cite{legetal2003}\cite{froetal2005}\cite{guzetal2005}\cite{guzetal2011},
where, using the so-called Fast Lyapunov Indicator (FLI; see
\cite{froetal2000}), a method was found to depict the resonant
structure of the action space in models of three degrees of freedom,
or 4D and 6D symplectic mappings being in the Nekhoroshev regime
\cite{froetal2000, froetal2005, guzetal2011}. In \cite{legetal2003},
the mean-square spread in action space $<\Delta J^2>$ was measured
as a function of the time $t$ for orbits along the chaotic border of
a {\it simply-resonant} domain (see section 2 for a precise
definition). It was found that i) the local character of diffusion
is normal, i.e. $<\Delta J^2>\propto t$, and ii) the diffusion
coefficient $D=<\Delta J^2>/t$ decreases with $\epsilon$ faster than
a power law. The exponential fit
$D\propto\exp(-(\epsilon_c/\epsilon)^{0.28}$ was given in a
subsequent study \cite{froetal2005}. The estimate obtained in
\cite{guzetal2011}, through interpolation over five orders of
magnitude of the perturbation parameter, yields with with certainty
the first digit of the exponent $0.2\ldots$, but the errors in the
interpolation make uncertain the second digit in both the above
estimates. In \cite{legetal2010a}, $D$ was measured as a function of
the separatrix splitting $S$ of the asymptotic manifolds of simply
unstable two-dimensional tori lying at the borders of simple
resonances (see also \cite{morgio1997}\cite{legetal2009}). The
measurement of $S$ itself was based on employing the FLI. It was
found that $D\propto S^p$, with $p=2.1$ and $p=2.56$ in two
resonances of increasing order respectively. Finally, the laws of
diffusion in systems violating one or more necessary conditions of
the Nekhoroshev theorem were investigated in
\cite{guzetal2006}\cite{legetal2010b}, leading to a number of
interesting results regarding the dynamical consequences of such
violations.

The motivation for the present study stems primarily from the
results reported in refs \cite{legetal2003}\cite{froetal2005}
\cite{guzetal2005} \cite{legetal2010a}, and it can be described as
follows. The results obtained so far are very satisfactory from the
numerical point of view. They require, however, computations
involving large ensembles of orbits and integration times of the
order of billions, or even trillions of periods. On the other hand,
we can remark that, in principle, the analytical methods involved in
the main theories of chaotic diffusion lend themselves also
conveniently to getting quantitative predictions regarding the value
of the diffusion coefficient, or the scaling laws of diffusion, in
general, in the weakly chaotic regime. For such a goal, however, to
be accomplished, it is required that one should be able to carry on
expansions of certain quantities up to a very high order in the
small parameter $\epsilon$ (usually with the aid of a computer).
This fact is explicit in Nekhoroshev theory, where one needs to
reach an expansion order high enough for the asymptotic behavior of
the perturbation series to show up. This has been realized in
studies seeking to determine the {\it range} (in the small parameter
value) and/or the {\it conditions of applicability} of Nekhoroshev
theory, or, finally, the {\it domain of practical stability} for
motions in simple physical systems or models inspired mainly from
Solar System dynamics (see e.g. \cite{gioetal1989}\cite{celfer1996}
\cite{morguz1997}\cite{giosko1997}\cite{eft2005}\cite{eftsan2005}
\cite{lhoetal2008}\cite{pavguz2008}\cite{gioetal2009}). These
studies notwithstanding, the question of central interest in the
present paper, namely how to obtain relevant quantitative estimates
of the {\it local value of the diffusion coefficient D} in resonant
domains (of various multiplicities) of the action space via {{\it
high order expansions} of perturbation theory, remains, to our
knowledge, largely unexplored.

Regarding now this last question, it should be noted that the formal
analytical apparatus of Nekhoroshev theory, entailing the
construction of a {\it normal form} in local domains covering the
action space of systems of the form (\ref{hamgen}), aims to
transform the original Hamiltonian into one in new canonical
variables resuming the form $H_{transformed}=Z+R$, where $Z$, the
normal form, corresponds to a simple dynamics, while $R$, the
remainder, induces a perturbation to this dynamics. The so-called
`geometric part' of Nekhoroshev the theorem ensures that, despite
allowing in general for chaotic motions, the flow under a
multiply-resonant normal form {\it alone} would imply perpetual
confinement of all chaotic orbits in balls of radius
$O(\epsilon^{1/2})$ in the action space. Nevertheless, this picture
is altered due to the effects of the remainder which eventually
causes the orbits to diffuse away of their initial
$O(\epsilon^{1/2})$ domain. Now, via a sequence of hamiltonian
normalization steps we find that there is an optimal order at which
the size of the remainder becomes exponentially small in a power of
$1/\epsilon$. This, in turn, implies an exponentially small {\it
semi-analytic upper bound} of the value of the diffusion coefficient
$D$. Unfortunately, such a bound turns usually to be very
unrealistic, as it overestimates by a large factor the true value of
$D$ (or, equivalently, it underestimates the time of practical
stability). We are thus led to conclude that, whereas the remainder
$R$ constitutes a quantity of primary interest in quantitative
applications of Nekhoroshev theory, the precise relation between $R$
and $D$ is apparently very different from what upper bound estimates
would suggest. Instead, a detailed analysis of the {\it effects of
the remainder on dynamics} appears to be necessary in order to
formulate a more precise theory of the relation between $R$ and $D$.

In the sequel, we present such an analysis in systems of three
degrees of freedom. In this analysis, we still have to rely on an
assumption for which numerical indications are available, namely
that the local character of diffusion in sufficiently small domains
of the action space is `normal', that is, the mean square spread of
the actions of the chaotic orbits grows linearly with time (there
are indications that {\it global diffusion}, which concerns
ensembles or orbits diffusing in a substantial part of the Arnold
web over much longer timescales, could also be described as `normal'
(see \cite{guzetal2005}); however, the issue of the laws of global
diffusion can only be hoped to tackle after the laws of local
diffusion have been adequately understood). In the rest of our
analysis, we proceed by expressing all quantities of interest in
terms of the remainder function, which, in turn, is calculated in
concrete examples by a well-defined algebraic procedure. Finally, we
estimate via this analysis how $D$ depends on the size $||R_{opt}||$
of the remainder at the optimal normalization order. It should be
noted that the idea that the stability properties of the orbits in
nearly-integrable systems depend on the size of the optimal
remainder is not new, but it is one permeating nearly all forms of
canonical perturbation theory. The novel feature here, instead, is
to use $||R_{opt}||$ not as an upper bound for $D$,  but as a way to
estimate $D$ via examining the relation between the two quantities
as determined by independent numerical experiments.  One main
prediction is that this relation is altered according to the {\it
multiplicity of resonance conditions} holding in the action domain
of interest. More concretely, we predict that the diffusion
coefficient $D$ scales with $||R_{opt}||$ as a power-law $D\propto
||R_{opt}||^p$, where $p\simeq 2$ in doubly-resonant domains, while
$p=2(1+b)$ in simply resonant domains, for some constant $b>0$. A
combination of theoretical arguments found in
\cite{chi1979}\cite{cin2002}, together with quantitative estimates
on the relation between the size of the so-called separatrix
splitting (see subsection 2.3.2) and the normal form remainder given
in \cite{morgio1997}, suggest $b\simeq 0.5$, i.e. $p\simeq 3$ in
simply resonant domains. This agrees with the numerical results
obtained in a previous study \cite{eft2008}.

In \cite{eft2008}, a computer-algebraic program was written in order
to calculate the optimal normal form as well as the remainder function
$R_{opt}$ at the optimal mormalization order in a case of simple
resonance, employing the same Hamiltonian model as in \cite{legetal2003}.
This operation involved computing about $5\times 10^7$ Fourier coefficients,
at a truncation order in Fourier space as high as $K=44$. Comparing
the computed size of $||R_{opt}||$ versus available numerical data on
$D$ from \cite{legetal2003}, the scaling $D\propto ||R_{opt}||^{2.98}$
was found by numerical fitting. In the present paper, after presenting
some theoretical results, we make a similar numerical calculation as
in \cite{eft2008} but in the case of a double resonance. In order
to reach the optimal normalization, we had to extend all normal form
calculations up to the Fourier order $K=50$ ($8\times 10^7$ coefficients).
We thus determined the size of the optimal remainder $||R_{opt}||$
for many different values of the small parameter $\epsilon$. In the
same time, we computed the diffusion coefficient $D$ for the same
values of $\epsilon$ by a purely numerical procedure involving runs
of ensembles of chaotic orbits (see section 3). Finally, we made two
independent numerical comparisons of the relation between $D$ and
$||R_{opt}||$. The latter yield the power laws $D\propto
||R_{opt}||^{2.1}$ and $D\propto ||R_{opt}||^{2.3}$ respectively.
This essentially confirms that $p$ is close to 2 in doubly resonant
domains, albeit with a small noticeable difference even in this case,
which probably requires a more precise theory to interpret.

Besides the above computation, our analysis using high order normal
forms resulted in a relevant result regarding the possibility to
visualize how the phenomenon of Arnold diffusion proceeds locally,
within a doubly-resonant domain, by materializing the computation of
a convenient set of variables helping to this purpose, that were
proposed in the work \cite{bengal1986}. We note that numerical
evidence for Arnold diffusion of orbits entering from simple to
double resonances was presented in \cite{guzetal2005}. Here, we
provide a detailed topological description of this phenomenon. The
whole computation consists of: i) computing a set of resonant
canonical action-angle variables via a sequence of Lie canonical
transformations, ii) taking a 2D Poincar\'{e} surface of section of
the {\it doubly-resonant normal form dynamics} (which represents a
system of two degrees of freedom), and (more importantly) iii) using
the {\it energy $E_Z$ of the normal form} as the third variable,
showing the effect of Arnold diffusion. According to theory, the
value of $E_Z$ changes exponentially slowly in time due to the
effect of the remainder. In the sequel we refer to this phenomenon
as `drift', although in reality it means that a number of quantities
can be characterized as undergoing random walk during the whole
diffusion process. Besides setting the timescale of diffusion, the
drift can be viewed also as the source of a dynamical phenomenon,
namely the communication between chaotic domains that would be
otherwise isolated under the doubly-resonant normal form hamiltonian
flow. We show in a true example the excursion of a chaotic orbit
within the doubly-resonant domain as it appears in the above
proposed set of variables. We thus identify a sequence of chaotic
transitions of such an orbit from one resonant domain to another. In
fact, in each transition the orbit bypasses the barriers imposed by
normal form dynamics via a `third dimension', i.e. the slowly
drifting value of $E_Z$. We finally argue that, besides their
practical utility, such illustrations are also suggestive of the
geometric structure underlying the asymptotic manifolds of
lower-dimensional tori filling the phase space in the domain of a
double resonance. These manifolds are important, because, following
the spirit of Arnold's original work \cite{arn1964}, it has been
widely conjectured that their heteroclinic intersections constitute
a primary cause of Arnold diffusion. Of course, proving this fact
represents a well known important open problem of dynamical systems'
theory.

The structure of the paper is as follows: Section 2 presents
the theory, focusing on the normal form algorithm, multiply-resonant
dynamics, effect of the remainder, and, finally, on the relation
between $D$ and $||R_{opt}||$. We describe in some length all necessary
theoretical steps in order to render the paper as self-contained
as possible. Section 3 then passes to the numerical results.
We present i) the results from the normal form computer-algebraic
construction, ii) the visualization of Arnold diffusion using
appropriate variables based on the normal form computation,
iii) the numerical calculation of the diffusion coefficient
$D$, and, finally iv) the comparison of $D$ with $||R_{opt}||$.
Section 4 summarizes the main conclusions of the present
study.

\section{Theory}
Most statements made in subsections 2.1 and 2.2 below, regarding the
properties of the Hamiltonian models considered as well as the algorithm
by which we perform Hamiltonian normalization, are applicable to systems
of an arbitrary number of degrees of freedom. In order, however, to be
consistent with the rest of the paper, we use everywhere a notation
referring to systems of three degrees of freedom. On the other hand,
the analysis of subsection 2.3 applies to the study of the diffusion
in doubly or simply resonant domains. In systems of three degrees
of freedom, the latter represent the only possible multiplicities
of a resonance condition, while in systems of more than three degrees
of freedom there are also cases of intermediate resonance multiplicities
between one and the maximal. The latter's study, nevertheless, is well
beyond our present computational capacity, and thus it is left as an
open problem.

\subsection{Definitions}
We consider three degrees of freedom systems of the form (\ref{hamgen}),
where $H$ satisfies the following analyticity and convexity conditions:

i) {\it Analyticity:} $H$ is assumed to be an analytic function in a
complexified domain of its arguments. Namely, we assume that there is
an open domain ${\cal I}\subset\mathbf{R}^3$ and a positive number
$\rho$ such that for all points $I_*\equiv(I_{1*},I_{2*},I_{3*})\in{\cal I}$
and all complex quantities $I_i'\equiv I_i-I_{i*}$ satisfying the inequalities
$|I_i'|<\rho$, the function $H_0$ admits a convergent Taylor
expansion
\begin{equation}\label{h0exp}
H_0=H_{0*}+\omega_*\cdot I'
+ \sum_{i=1}^3\sum_{j=1}^3{1\over 2}M_{ij*}I_i'I_j' +\ldots
\end{equation}
where $\omega_*=\nabla_I H_0(I_*)$, and $M_{ij*}$ are the entries of the
Hessian matrix of $H_0$ at $I_*$. Furthermore, we assume that there is
a positive constant $\sigma$ such that for all $I\in {\cal I}$,
$H_1$ admits an absolutely convergent Fourier  expansion
\begin{equation}\label{h1four}
H_1=\sum_k h_k(I)\exp(ik\cdot\phi)
\end{equation}
in a domain where all three angles satisfy $0\leq Re(\phi_i)<2\pi$,
$|Im(\phi_i)|\leq \sigma$. By the Fourier theorem (see e.g. \cite{gio2002}),
this condition implies that the coefficients $h_k(I)$ decay
exponentially with the $L^1$--modulus $|k|\equiv|k_1|+|k_2|+|k_3|$,
that is, there is a positive constant $A$ such that the bound
\begin{equation}\label{anal}
|h_k(I)|< A\exp(-|k|\sigma)
\end{equation}
holds for all $k\in {\cal Z}^3$. We finally assume that all
coefficients $h_k$ admit Taylor expansions with respect to $I_*$
\begin{equation}\label{hkexp}
h_k=h_{k*}+\nabla_{I_*}h_k\cdot I'
+ {1\over 2}\sum_{i=1}^3\sum_{j=1}^3h_{k,ij*}I_i'I_j' +\ldots
\end{equation}
(where $h_{k,ij*}$ are the entries of the Hessian matrix of $h_k(I)$
at $I_*$), which are convergent in the same union of domains as for
$H_0$.

ii) {\it Convexity:} For the Hessian matrix $M_*$, which is real symmetric,
we assume a simple quasi-convexity condition, namely that for all
$I_*\in{\cal I}$ either two of the (real) eigenvalues of $M_*$ have
the same sign and one is equal to zero, or all three eigenvalues
have the same sign. Furthermore, we define two constants:
\begin{equation}\label{mh}
\mu_{min}=\min\{|\mu_j|\},~~~\mu_{max}=\max\{|\mu_j|\}
\end{equation}
where $j$ is a label of only non-zero eigenvalues $\mu_j$ of $M_*$,
i.e. $j=1,2$ or $j=1,2,3$ if there are two or three non-zero eigenvalues
respectively.

As will be discussed in detail in subsection 2.3, the
quasi-convexity condition is essential, since it introduces a
confinement of the orbits for exponentially long times on a surface
arising from the condition of preservation of the energy (see
\cite{bengal1986}).

We now give some definitions allowing to characterize resonant dynamics.

A {\it resonant manifold} ${\cal R}_k$ associated with a non-zero
wavevector $k$ with co-prime integer components $k\equiv(k_1,k_2,k_3)$
is the two-dimensional locus defined by
\begin{equation}\label{resman}
{\cal R}_k=\left\{I\in{\cal I}:
k_1\omega_1(I)+k_2\omega_2(I)+k_3\omega_3(I)=0\right\}~~~,
\end{equation}
where $\omega_i(I)=\partial H_0/\partial I_i$.

Let $I_*\in{\cal I}$ be such that all three frequencies $\omega_i(I_*)$,
$i=1,2,3$ are different from zero. We now distinguish the following
three cases:\\

i) {\it Non-resonance:} no resonant manifold ${\cal R}_k$ contains $I_*$.

ii) {\it Simple resonance:} one resonant manifold ${\cal R}_k$ contains
$I_*$.

iii) {\it Double resonance:} more than one resonant manifolds contain
$I_*$. In the latter case, it is possible to choose two
linearly independent vectors $k^{(1)},k^{(2)}$ such that all resonant
manifolds ${\cal R}_{k}$ containing $I_*$ are labeled by vectors $k$
which are linear combinations of the chosen vectors $k^{(1)},k^{(2)}$
with rational coefficients. The intersection of these manifolds forms
a one-dimensional {\it resonant junction}. A doubly-resonant point
$I_*$ always corresponds to the intersection of a resonant junction
with a constant energy surface $H_0(I_*)=E$. \\

In the above definitions, resonant manifolds ${\cal R}_k$ of all
possible wavevectors $k$ have been considered. It is well known,
however, that in normal form theory a natural truncation limit
$|k|<K$ arises in Fourier space (see below). Accounting for this
possibility, we call a point $I_*\in{\cal I}$ i) non-resonant,
ii) simply resonant, or iii) doubly resonant {\it with respect to
a K--truncation}, if the number of resonant manifolds ${\cal R}_k$
with $|k|<K$ passing through $I_*$ are i) zero, ii) one and iii) more
than one respectively.

Finally, it will be convenient to introduce a definition concerning
{\it open domains} in ${\cal I}$. Let ${\cal W}_{I_*,B}$ be a ball
of radius $B$ around one point $I_*$ in ${\cal I}$. If $H_0$ satisfies
convexity conditions as assumed above, for $B$ small whatsoever the
domain ${\cal W}_{I_*,B}$ is crossed by a dense set of resonant manifolds
${\cal R}_k$. However, for any fixed value of the positive integer
$K$, only a finite subset of the manifolds ${\cal R}_k$ satisfy
$|k|<K$. The domain ${\cal W}_{I_*,B}$ is then called:
i) non-resonant, ii) simply-resonant, and iii) doubly-resonant
with respect to the $K$--truncation if $I_*$ is, respectively,
non-resonant, simply-resonant or doubly-resonant, and no other
resonant manifolds ${\cal R}_k$ with $|k|<K$ cross
${\cal W}_{I_*,B}$ except for the ones passing through $I_*$.

\subsection{Normal form construction}
All our estimates on the speed of diffusion are based on an appropriate
normal form construction. In this, we adopt the method exposed in detail
in \cite{eft2008}, which lends itself conveniently to i) developing a
computer-algebraic program, and ii) deriving analytical estimates on
the size of various quantities appearing in the course of Hamiltonian
normalization. The main elements of this method are:\\

{\it Expansion centers.} The action space can be covered by domains
${\cal W}_{I_*,B}$, centered around points $I_*$ which serve as
expansion centers of both the original Hamiltonian and the normal
form. We choose the points $I_*$ to belong to the set of all
doubly-resonant points of ${\cal I}$, denoted by  ${\cal D}$, and by
setting $B$ as of order $O(\epsilon^{1/2})$. The covering is possible
because ${\cal D}$ is dense in ${\cal I}$. A normal form construction
as done below is valid within one domain ${\cal W}_{I_*,B}$ (this
is essentially the same starting point as in Lochak's \cite{loch1992}
analytic construction leading to a proof of the Nekhoroshev theorem).
A crucial remark is that the characterization of dynamics within
${\cal W}_{I_*,B}$ as non resonant, simply resonant, or doubly
resonant depends on $\epsilon$. This is because, as shown below,
the optimal normal form truncation order $K=K_{opt}$ in Fourier
space depends on the value of $\epsilon$. Furthermore, for a given
value of $K$, the set ${\cal D}$ can be decomposed in three disjoint
sets ${\cal D}={\cal D}_{0,K}\cup{\cal D}_{1,K}\cup{\cal D}_{2,K}$,
containing all non-resonant, simply resonant and doubly resonant points
respectively with respect to the $K$--truncation. Thus, the characterization
of resonant dynamics within ${\cal W}_{I_*,B}$ depends on whether,
according to the value of $K$, $I_*$ belongs to ${\cal D}_{0,K}$,
${\cal D}_{1,K}$, or ${\cal D}_{2,K}$.\\

{\it Resonant module:} Let $I_*$ be a point of ${\cal D}$ and
$k^{(1)}\equiv(k^{(1)}_1,k^{(1)}_2,k^{(1)}_3)$, $k^{(2)}
\equiv(k^{(2)}_1,k^{(2)}_2,k^{(2)}_3)$ two linearly independent vectors
such that $k^{(i)}\cdot\omega(I_*)=0$ for $i=1,2$.  More than one
choices of $k^{(1)}$ and $k^{(2)}$ are possible. In the sequel we
choose $k^{(1)}$ and $k^{(2)}$ so that $|k^{(1)}|+|k^{(2)}|$ is
minimal. The vector $m\equiv(m_1,m_2,m_3)$ defined by
\begin{equation}\label{mvec}
m_1=k^{(1)}_2k^{(2)}_3-k^{(2)}_2k^{(1)}_3,~~~
m_2=k^{(1)}_3k^{(2)}_1-k^{(2)}_3k^{(1)}_1,~~~
m_3=k^{(1)}_1k^{(2)}_2-k^{(2)}_1k^{(1)}_2~~~
\end{equation}
is parallel to the vector $\omega(I_*)$ since $k\cdot m=0$ for all
$k$ satisfying $k\cdot\omega(I_*)=0$. If $m_1$, $m_2$, $m_3$
are not co-prime integers, we re-define $m$ by dividing the
$m_i$ by their greatest common divisor. The set
\begin{equation}\label{resmod}
{\cal M}\equiv\left\{k\in {\cal Z}^3: k\cdot m=0\right\}
\end{equation}
is hereafter called the resonant module associated with the point
$I_*\in{\cal D}$. The resonant module includes wavevectors $k$ whose
respective trigonometric terms $\exp(ik\cdot\phi)$ are to be retained
in the normal form. \\

{\it Action re-scaling:} From now on we focus on the construction of the normal
form in one specific domain ${\cal W}_{I_*,B}$. It has been mentioned already
that it is convenient to choose $B$ as a quantity scaling proportionally to
$\epsilon^{1/2}$. The simplest way to accommodate such a choice is by
introducing the following re-scaling of all action variables within
${\cal W}_{I_*,B}$:
\begin{equation}\label{resc}
J_i=\epsilon^{-1/2}(I_i-I_{i*})=\epsilon^{-1/2} I_i',~~~i=1,2,3~~.
\end{equation}
This re-scaling greatly simplifies the normal form algorithm, because it
formally removes all terms besides linear in the actions from the kernel
of the so-called homological equation (see below, or \cite{eft2008} for
details) by which the normalizing generating functions are determined.
Eq.(\ref{resc}) does not define a canonical transformation. However,
the correct equations of motion in the variables $(J,\phi)$ are produced
by the Hamiltonian function $h(J,\phi)=\epsilon^{-1/2}
H(I_*+\epsilon^{1/2}J,\phi)$, i.e. (neglecting a constant)
\begin{eqnarray}\label{hamexp2}
h(J,\phi)&=&\omega_*\cdot J
+ \epsilon^{1/2}\sum_{i=1}^3\sum_{j=1}^3{1\over 2}M_{ij*}J_iJ_j +\ldots\\
&+&\epsilon^{1/2}\sum_k \left(h_{k*}+\epsilon^{1/2}\nabla_{I_*}h_k\cdot J
+ {\epsilon\over 2}\sum_{i=1}^3\sum_{j=1}^3h_{k,ij*}J_iJ_j +\ldots\right)
\exp(ik\cdot\phi)\nonumber
\end{eqnarray}
where the first line in the above equation comes from the integrable
part $H_0$ of the original Hamiltonian (Eq.(\ref{h0exp})), while the
second line comes from the perturbation $H_1$ (Eq.(\ref{h1four}))
given the series expansion of the Fourier coefficients as in
Eq.(\ref{hkexp}).\\

{\it Book-keeping:} We now split the Hamiltonian (\ref{hamexp2}) in
parts of different order of smallness, which are to be normalized step
by step. The function (\ref{hamexp2}) contains terms of various orders
in the small parameter $\epsilon^{1/2}$. However, the presence
of a second `small parameter' $e^{-\sigma}$ is implied in (\ref{hamexp2})
by the exponential decay of all Fourier coefficients $h_{k*}$, $h_{k,ij*}$,
etc., due to Eq.(\ref{anal}) (see \cite{gio2002} pp.90-91 for a thorough
exposition of the role of this small parameter in Nekhoroshev theory).
We take both parameters into account by introducing an integer $K'$
such that $e^{-\sigma K'}\sim\epsilon^{1/2}$, i.e. by setting:
\begin{equation}\label{kprime}
K'=\left[-{1\over 2\sigma}\ln(\epsilon)\right]~~.
\end{equation}
Using $K'$, the Hamiltonian (\ref{hamexp2}) can be split in groups
of practically the same order of smallness. This is realized by
artificially introducing a `book-keeping' coefficient $\lambda^p$
in front of each term in (\ref{hamexp2}), whose numerical value is
set equal to unity at the end of the calculation. Furthermore, for a
term of the form $\epsilon^{\mu/2}f(J)\exp(ik\cdot\phi)$ we set
$p=[|k|/K']+\mu$.

Regarding the above `book-keeping' process it is worth noting the
following: i) This way of splitting the Hamiltonian in different
orders of smallness results in a finite number of terms appearing in
every power of $\lambda$. ii) This technique is suggested already by
Poincar\'{e} \cite{poi1892} and Arnold \cite{arn1963}. In fact, the
dependence of $K'$ on $\epsilon$ is weak, since it is logarithmic,
so that an alternative choice to the `ansatz' (\ref{kprime}) is to
set $K'=const\sim 1/\sigma$. In fact, according to Giorgilli
\cite{gio2002} this is an optimal choice. iii) Since, at every
normalization order, we have a reduction of the analyticity domain,
one could consider re-defining $K'$ at every normalization step.
However, this is hardly tractable from an algorithmic point of view.
Instead, keeping $K'$ constant at all normalization orders should be
viewed as a rule indicating the sequence by which the various terms
in the Hamiltonian are normalized, i.e., the terms or order
$\lambda^r$ are normalized in the r-th step. Albeit not necessarily
optimal regarding the grouping of the terms according to their size,
this rule proves simple to implement and sufficient in practice.

Returning to the form of the Hamiltonian after introducing the
book-keeping factor} $\lambda$, the Hamiltonian reads:
\begin{eqnarray}\label{hamexp3}
h&=&\omega_*\cdot J
+ \lambda\epsilon^{1/2}
   \sum_{i=1}^3\sum_{j=1}^3{1\over 2}M_{ij*}J_iJ_j +\ldots
+\sum_k \Bigg(\lambda^{1+[|k|/K']}\epsilon^{1/2}h_{k*}\\
&+&\lambda^{2+[|k|/K']}\epsilon\nabla_{I_*}h_k\cdot J
+ \lambda^{3+[|k|/K']}{\epsilon^{3/2}\over 2}
\sum_{i=1}^3\sum_{j=1}^3h_{k,ij*}J_iJ_j +\ldots\Bigg)
\exp(ik\cdot\phi)\nonumber~~.
\end{eqnarray}\\
Setting $Z_0=\omega_*\cdot J$, the Hamiltonian (\ref{hamexp3})
resumes the form
\begin{eqnarray}\label{hamexpf}
h=H^{(0)}(J,\phi)&=&Z_0+\sum_{s=1}^{\infty}\lambda^s
H^{(0)}_s(J,\phi;\epsilon^{1/2})
\end{eqnarray}
where i) the superscript $(0)$ denotes zeroth-step of the normalization
procedure (= original Hamiltonian), ii) the exponent of $\lambda$ in
different terms keeps track of their  true order of smallness, and
iii) the functions $H^{(0)}_s$ are of the form
\begin{equation}\label{h0s}
H^{(0)}_s = \sum_{\mu=1}^s\epsilon^{\mu/2}
\sum_{|k|=K'(s-\mu)}^{K'(s-\mu+1)-1} H^{(0)}_{\mu,k}(J)\exp(ik\cdot\phi)
\end{equation}
where $H^{(0)}_{\mu,k}(J)$ are polynomials containing terms of degree
$\mu-1$ or $\mu$ in the action variables $J$. Precisely, we have:
$$
H^{(0)}_{\mu,k}(J)=
\sum_{\mu_1=0}^{\mu-1}~~
\sum_{\mu_2=0}^{\mu-1-\mu_1}~~
\sum_{\mu_3=0}^{\mu-1-\mu_1-\mu_2}
{1\over\mu_1!\mu_2!\mu_3!}
{\partial^{\mu-1}h_{k}(I_*)\over
\partial^{\mu_1}I_1\partial^{\mu_2}I_2\partial^{\mu_3}I_3}
J_1^{\mu_1}J_2^{\mu_2}J_3^{\mu_3}
$$
if $|k|>0$, or
$$
H^{(0)}_{\mu,k}(J)=
\sum_{\mu_1=0}^{\mu}~~
\sum_{\mu_2=0}^{\mu-\mu_1}~~
\sum_{\mu_3=0}^{\mu-\mu_1-\mu_2}
{1\over\mu_1!\mu_2!\mu_3!}
{\partial^{\mu}H_0(I_*)\over
\partial^{\mu_1}I_1\partial^{\mu_2}I_2\partial^{\mu_3}I_3}
J_1^{\mu_1}J_2^{\mu_2}J_3^{\mu_3}
$$
$$
~~~~~~~~+\sum_{\mu_1=0}^{\mu-1}~~
\sum_{\mu_2=0}^{\mu-1-\mu_1}~~
\sum_{\mu_3=0}^{\mu-1-\mu_1-\mu_2}
{1\over\mu_1!\mu_2!\mu_3!}
{\partial^{\mu-1}h_{0}(I_*)\over
\partial^{\mu_1}I_1\partial^{\mu_2}I_2\partial^{\mu_3}I_3}
J_1^{\mu_1}J_2^{\mu_2}J_3^{\mu_3}
$$
if $k=0$.\\

{\it Hamiltonian normalization:} We use the algorithm of composition of Lie
series in order to perform the Hamiltonian normalization. Let us recall that
the purpose of the normalization is to introduce a sequence of canonical
transformations $(J,\phi)\equiv (J^{(0)},\phi^{(0)})$
$\rightarrow(J^{(1)},\phi^{(1)})$ $\rightarrow(J^{(2)},\phi^{(2)})\rightarrow\ldots$
so that the Hamiltonian expressed as a function of the new variables allows
one to more easily identify the main features of dynamics. After $r$
normalization steps, the old variables $(J,\phi)\equiv(J^{(0)},\phi^{(0)})$
are expressed in terms of the new variables $(J^{(r)},\phi^{(r)})$, and the
Hamiltonian $H^{(r)}(J^{(r)},\phi^{(r)})=h(J(J^{(r)},\phi^{(r)}),
\phi(J^{(r)},\phi^{(r)}))$ takes the form
\begin{equation}\label{nfrem}
H^{(r)}(J^{(r)},\phi^{(r)})=Z^{(r)}(J^{(r)},\phi^{(r)};\lambda,\epsilon)
+R^{(r)}(J^{(r)},\phi^{(r)};\lambda,\epsilon)~~~.
\end{equation}
The terms $Z^{(r)}(J^{(r)},\phi^{(r)};\lambda,\epsilon)$ and
$R^{(r)}(J^{(r)},\phi^{(r)};\lambda,\epsilon)$ are called the {\it normal
form} and the {\it remainder} respectively. The normal form is a finite expression
which contains terms up to order $r$ in the book-keeping constant $\lambda$, while
the remainder is a series containing terms of order $\lambda^{r+1}$ and beyond.
The mathematical structure of the normal form term $Z^{(r)}$ is such as
to imply an easily identifiable dynamics in the variables $(J^{(r)}
,\phi^{(r)})$ (e.g. an oscillator or pendulum dynamics). On the other hand,
the remainder is a {\it convergent} series in a restriction of the domain
of analyticity of the original Hamiltonian, which represents a perturbation
with respect to the Hamiltonian flow of $Z^{(r)}$. An optimal normalization
order $r_{opt}$ exists (see below) where the process must be stopped.

The Hamiltonian normalization is implemented step-by-step by the recursive
equation:
\begin{equation}\label{hr}
H^{(r)} = \exp(L_{\chi_r})H^{(r-1)}
\end{equation}
where $\chi_r$ is the r-th step Lie generating function and $L_{\chi_r}\equiv
\{\cdot,\chi_r\}$ is the Poisson bracket operator. Both $H^{(r)}$ and $\chi_r$
are functions of the variables $J^{(r)},\phi^{(r)}$. The generating function
is defined by the solution of the homological equation
\begin{equation}\label{homo}
\{\omega_*\cdot J^{(r)},\chi_r\}+\tilde{H}^{(r-1)}_r(J^{(r)},\phi^{(r)})=0
\end{equation}
where $\tilde{H}^{(r-1)}_r(J^{(r)},\phi^{(r)})$ denotes all terms of $H^{(r-1)}$
which i) have a book-keeping coefficient $\lambda^r$ in front, and ii) belong
to the range of the operator $\{\omega_*\cdot J^{(r)},\cdot\}$. Given the
definition of the resonant module ${\cal M}$ in Eq.(\ref{resmod}), one has
the relation
\begin{equation}\label{hrtilde}
\tilde{H}^{(r-1)}_r = H^{(r-1)}_r - Z_r
\end{equation}
where $H^{(r-1)}_r$ are all the terms of $H^{(r-1)}$ having a factor
$\lambda^r$, and $Z_r$ are the {\it normal form} terms of $H^{(r-1)}_r$,
that is all the trigonometric terms whose wavevectors $k$ belong to ${\cal M}$.
It follows immediately that $H^{(r)}$ has the form
\begin{equation}\label{hamnf}
H^{(r)}=Z_0+Z_1+...+Z_r+ R^{(r)}
\end{equation}
where all terms in the functions $Z_i$ have a factor $\lambda^i$, while
$R^{(r)}$ is a series in powers of $\lambda$ starting with terms of order
$\lambda^{r+1}$. \\

{\it Optimal truncation:}  In the analytical part of the Nekhoroshev theory it
is demonstrated that the whole normalization process has an asymptotic character.
Namely, i) the domain of convergence of the remainder series $R^{(r)}$ shrinks
as the normalization order $r$ increases, and ii) the size $||R^{(r)}||$ of
$R^{(r)}$, where $||\cdot||$ is a properly defined norm in the space of
trigonometric polynomials (see below), initially decreases, as $r$ increases,
up to an optimal order $r_{opt}$ beyond which $||R^{(r)}||$ increases with $r$.
In the Nekhoroshev regime, one has $||Z^{(r_{opt})}||>>||R^{(r_{opt})}||$.
Thus, stopping at $r_{opt}$ best unravels the dynamics, which is given
essentially by the Hamiltonian flow of $Z^{(r_{opt})}$ slightly perturbed
by $R^{(r_{opt})}$. The long term consequences of this perturbation,
which determine the speed of diffusion, will be analyzed in subsection
2.3.

The normal form $Z^{(r)}=Z_0+Z_1+...Z_r$ contains trigonometric
terms $\exp(ik\cdot\phi)$ of order not greater than $K=K'r-1$. Let
$r_{opt}$ be the optimal normalization order. It is well known that
the dependence of $r_{opt}$ on $\epsilon$ is given by an inverse
power-law, namely
\begin{equation}\label{ropt}
r_{opt}\sim\epsilon^{-a}~~.
\end{equation}
The exponents $1/6$, $1/4$ and $1/2$, referring to the non-resonant,
simply resonant, and doubly resonant normal form constructions
respectively, are found in \cite{posh1993}. We emphasize that,
while, due to the introduction of the book-keeping process, the
algorithm of Hamiltonian normalization analyzed above is not
technically identical with the usual normalization procedure used in
the proof of the Nekhoroshev theorem (e.g. as in \cite{posh1993}),
in practice we recover the estimate (\ref{ropt}), and the resulting
exponents, both in the simply resonant case (see \cite{eft2008}) and
in the doubly resonant case, as confirmed by numerical experiments
in section 3 below. In particular, we find that since the leading
terms in the remainder are $O(\lambda^{r_{opt}+1})$, the size of the
remainder is of order $O(\epsilon^{(r_{opt}+1)/2})$, implying
(viz.Eq.(\ref{kprime})):
\begin{equation}\label{remropt}
||R^{(r_{opt})}||\sim \epsilon^{1/2}\exp\left({-K'\sigma\over\epsilon^{a}}\right)
\end{equation}
i.e. the remainder is exponentially small in $1/\epsilon$ in accordance with
the Nekhoroshev theorem. The Fourier order
\begin{equation}\label{kopt}
K_{opt}(\epsilon)=K'r_{opt}(\epsilon)
\end{equation}
is hereafter called the optimal K--truncation order. All the normal form terms
of $H^{(r_{opt})}$ have Fourier orders satisfying $|k|<K_{opt}(\epsilon)$.\\

\subsection{Resonant normal form dynamics and the rate of diffusion}
We are now in a position to discuss the essence of all the previous
definitions. The key point is to observe that, depending on the
value of $\epsilon$, {\it the same} expansion point $I_*\in{\cal D}$
of the normal form construction turns to be either non-resonant, or
simply or doubly resonant with respect to the optimal K--truncation.
In particular, let $k^{(1)}$ and $k^{(2)}$ be two linearly
independent vectors of ${\cal M}$ such that for all $k\in{\cal M}$
one has $|k|\geq |k^{(2)}|\geq |k^{(1)}|$. We then distinguish the
following three regimes: i) $|k^{(2)}|< K_{opt}(\epsilon)$. Then,
the point $I_*$ is doubly-resonant with respect to the optimal
K--truncation. This is the case we mainly focus on in the sequel.
The main theoretical results are given in subsection 2.3.1, while
the main numerical results are given in section 3. ii)
$|k^{(1)}|<K_{opt}(\epsilon)\leq |k^{(2)}|$. Then, $I_*$ is
simply-resonant with respect to the optimal K--truncation. One such
example was dealt with in the numerical study \cite{eft2008}.
Further theoretical analysis of this case is made in subsection
2.3.2. iii) $K_{opt}(\epsilon)< |k^{(1)}|\leq |k^{(2)}|$. Then,
$I_*$ is non-resonant with respect to the optimal K--truncation.
Since $K_{opt}$ decreases as $\epsilon$ increases, for fixed
$|k^{(1)}|+|k^{(2)}|$ this inequality always occurs if
$\epsilon>\epsilon_1$, where $\epsilon_1$ is a threshold depending
on $k^{(1)}$, $k^{(2)}$. The case $\epsilon_1>\epsilon_c$, where
$\epsilon_c$ is the critical threshold for the onset of the
Nekhoroshev regime, presents no practical interest. If, however,
$\epsilon_1<\epsilon_c$, then, for all values of $\epsilon$ in the
interval $\epsilon_1<\epsilon<\epsilon_c$ the optimal normal form
describes a true non-resonant dynamics. Note that in order to
describe the dynamics close to a point $I_*'$ of the action space
corresponding to Diophantine frequencies $\omega_*'$, it suffices to
choose $I_*$ such that $\omega_*$ corresponds to a very high order
rational approximation of $\omega_*$, i.e. the numbers
$(\omega_{1*},\omega_{2*}, \omega_{3*})$ are high order finite digit
approximants of the numbers
$(\omega_{1*}',\omega_{2*}',\omega_{3*}')$. Then,
$|k^{(1)}|+|k^{(2)}|$ becomes very large, and $\epsilon_1$
approaches very close to zero. In this case, for $\epsilon$
sufficiently small, we expect the existence of a set of points of
large measure within ${\cal W}_{I_*,B}$, corresponding to Kolmogorov
- Arnold - Moser tori in the neighborhood of the point $I_*$.
However, these tori cannot fill an open domain. Thus, the diffusion
in action space is topologically possible for (very weakly) chaotic
orbits wandering through the set of KAM tori. However, in the
absence of significant resonant chaotic layers (since no important
resonances cross ${\cal W}_{I_*,B}$), the question of whether or not
the diffusion can be observed is of no practical interest, since its
rate would be extremely slow to be of any relevance in applications.
Thus, the non-resonant case is no further considered below.

\subsubsection{Double resonance}
As long as $|k^{(2)}|<K_{opt}(\epsilon)$, the point $I_*$ is
doubly-resonant with respect to the optimal K--truncation. In this case,
the normal form contains either terms independent of the angles,
or trigonometric terms of the form $\exp(ik^{(1)}\cdot\phi^{(r_{opt})})$,
$\exp(ik^{(2)}\cdot\phi^{(r_{opt})})$ and their multiples in the exponents.
Writing explicitly only the most important terms, the normalized Hamiltonian
takes the form:
\begin{eqnarray}\label{hamres}
h(J^{(r_{opt})},\phi^{(r_{opt})})&=&Z(J^{(r_{opt})}
,\phi^{(r_{opt})})+R(J^{(r_{opt})},\phi^{(r_{opt})})\nonumber\\
&=&\omega_*\cdot J^{(r_{opt})}
+ \epsilon^{1/2}
\sum_{i=1}^3\sum_{j=1}^3{1\over 2}M_{ij*}J^{(r_{opt})}_iJ^{(r_{opt})}_j
+\ldots\\
&+&\epsilon^{1/2}\sum_{n_1,n_2\in Z^2}g_{n_1,n_2}(J^{(r_{opt})})
\exp(i(n_1k^{(1)}+n_2k^{(2)})\cdot\phi^{(r_{opt})})+\ldots\nonumber\\
&+&R(J^{(r_{opt})},\phi^{(r_{opt})}))\nonumber
\end{eqnarray}
The main feature of the Hamiltonian (\ref{hamres}) is that, since in
$Z(J^{(r_{opt})},\phi^{(r_{opt})}))$ there are coupling terms
between more than one resonant angles, the normal form $Z$ {\it
alone} is {\it non-integrable}. In fact, $Z$ can be decomposed into
an integrable system of one degree of freedom and a non-integrable
system of two degrees of freedom  (see \cite{bengal1986}). The
decomposition is done by the linear canonical transformation
$(J^{(r_{opt})}_1,J^{(r_{opt})}_2,J^{(r_{opt})}_3,
\phi^{(r_{opt})}_1,\phi^{(r_{opt})}_2,\phi^{(r_{opt})}_3)$
$\rightarrow$ $(J_{R_1},J_{R_2},J_F,\phi_{R_1},\phi_{R_2},\phi_F)$
defined by
\begin{eqnarray}\label{restra}
J^{(r_{opt})}_1&=k_1^{(1)}J_{R_1}+k_1^{(2)}J_{R_2}+m_1J_F,~~~~~
\phi_{R_1}&=k_1^{(1)}\phi^{(r_{opt})}_1+k_2^{(1)}\phi^{(r_{opt})}_2
+k_3^{(1)}\phi^{(r_{opt})}_3\nonumber\\
J^{(r_{opt})}_2&=k_2^{(1)}J_{R_1}+k_2^{(2)}J_{R_2}+m_2J_F,~~~~~
\phi_{R_2}&=k_1^{(2)}\phi^{(r_{opt})}_1+k_2^{(2)}\phi^{(r_{opt})}_2
+k_3^{(2)}\phi^{(r_{opt})}_3\\
J^{(r_{opt})}_3&=k_3^{(1)}J_{R_1}+k_3^{(2)}J_{R_2}+m_3J_F,~~~~~
\phi_F&=m_1\phi^{(r_{opt})}_1+m_2\phi^{(r_{opt})}_2
+m_3\phi^{(r_{opt})}_3\nonumber
\end{eqnarray}
where $m\equiv(m_1,m_2,m_3)$ has been defined in Eq.(\ref{mvec}).
The Hamiltonian in the new variables reads (apart from a constant)
\begin{eqnarray}\label{hamres2}
&h=
Z(J_{R_1},J_{R_2},J_F,\phi_{R_1},\phi_{R_2})+
R_{opt}(J_{R_1},J_{R_2},J_F,\phi_{R_1},\phi_{R_2},\phi_F)
\end{eqnarray}
where
\begin{eqnarray}\label{nfdble}
&~&Z(J_{R_1},J_{R_2},J_F,\phi_{R_1},\phi_{R_2})
=(\omega_*\cdot m) J_F\nonumber\\
&~&+\epsilon^{1/2}\sum_{i,j=1}^3{1\over 2}
M_{ij*}(k_i^{(1)}J_{R_1}+k_i^{(2)}J_{R_2}+m_iJ_F)
(k_j^{(1)}J_{R_1}+k_j^{(2)}J_{R_2}+m_jJ_F) +\ldots\\
&~&+\epsilon^{1/2}\sum_{n_1,n_2\in Z^2}g_{n_1,n_2}(J_{R_1},J_{R_2},J_F)
\exp(i(n_1\phi_{R_1}+n_2\phi_{R_2}))+\ldots\nonumber
\end{eqnarray}
and the remainder
$R_{opt}(J_{R_1},J_{R_2},J_F,\phi_{R_1},\phi_{R_2},\phi_F)$
is exponentially small in $1/\epsilon$. Since $\phi_F$ is ignorable
in $Z$, $J_F$ is an integral under the flow of the normal form.
On the other hand, the remaining degrees of freedom
$(J_{R_1},\phi_{R_1})$ and $(J_{R_2},\phi_{R_2})$ are
coupled under the flow of $Z$ due to the trigonometric terms
$\exp(i(n_1\phi_{R_1}+n_2\phi_{R_2}))$.
The main characteristics of motion can be understood by the following remarks
\footnote{Since many different action symbols appear in the previous and
in the subsequent analysis, it helps recalling that throughout the paper
all action variables defined by a symbol starting with the letter $I$
refer to non-scaled values, i.e. before the re-scaling of Eq.(\ref{resc})
is implemented, while all action variables defined by a symbol starting
with the letter $J$ have re-scaled values, according to Eq.(\ref{resc}).
Thus, in the domains considered below, all quantities of the form
$I-I_*$, where $I_*$ is the selected central doubly-resonant point
of interest, scale proportionally to $\epsilon^{1/2}$, while all
actions denoted by a letter $J$ exhibit no scaling with $\epsilon$.
Furthermore, all Hamiltonian-type functions denoted by $h$, $H^{(r)}$,
$Z$, or $R$, are expressed in re-scaled variables; only the original
Hamiltonian (Eq.(\ref{hamgen})) is expressed in non-scaled action variables
$I$. Finally, the quantities $E'$ (Eq.(\ref{eneprime})) and $E_Z$
(Eq.(\ref{ez})) scale proportionally to $\epsilon^{1/2}$.}:

i) The constant-valued action $J_F$ can be viewed as a parameter in the
two degrees of freedom Hamiltonian $Z$. Furthermore, except for the
case of some very low resonances satisfying $|k^{(1)}|<K'$, all
coefficients $g_{n_1,n_2}$ in (\ref{nfdble}) are of order
$\epsilon^{1/2}$ or higher. Thus, the terms
\begin{eqnarray}\label{nfdble0}
&~&Z_0(J_{R_1},J_{R_2};J_F)=(\omega_*\cdot m) J_F\nonumber\\
&~&+\epsilon^{1/2}\sum_{i,j=1}^3{1\over 2}
M_{ij*}(k_i^{(1)}J_{R_1}+k_i^{(2)}J_{R_2}+m_iJ_F)
(k_j^{(1)}J_{R_1}+k_j^{(2)}J_{R_2}+m_jJ_F) +\ldots
\end{eqnarray}
define an `integrable part' of $Z$, while the remaining terms
depending on the resonant angles can be considered as a perturbation.

The terms quadratic in $J_{R_1},J_{R_2}$ in the r.h.s. of (\ref{nfdble0})
define the quadratic form
\begin{equation}\label{z02}
\zeta_{0,2}={1\over 2}
\sum_{i,j=1}^3
M_{ij*}(k_i^{(1)}J_{R_1}+k_i^{(2)}J_{R_2})
(k_j^{(1)}J_{R_1}+k_j^{(2)}J_{R_2})
\end{equation}
In Appendix A it is demonstrated that, due to the quasi-convexity condition
assumed for the Hessian matrix $M_{ij*}$, the quadratic form (\ref{z02}) is
positive definite. Thus, the constant level curves of the quantity
\begin{equation}\label{eneprime}
E'=(Z_0-(\omega_*\cdot m) J_F)
\end{equation}
on the plane $(J_{R_1},J_{R_2})$, given by
\begin{eqnarray}\label{ene0}
E'=\epsilon^{1/2}\sum_{i,j=1}^3{1\over 2}
M_{ij*}(k_i^{(1)}J_{R_1}+k_i^{(2)}J_{R_2}+m_iJ_F)
(k_j^{(1)}J_{R_1}+k_j^{(2)}J_{R_2}+m_jJ_F)~~,
\end{eqnarray}
are ellipses centered at
\begin{eqnarray}\label{jr0}
J_{R_1,0}&=&\frac
{\left(k^{(1)}\cdot M_*k^{(2)}\right) \left(m\cdot M_*k^{(2)}\right)
-\left(k^{(2)}\cdot M_*k^{(2)}\right) \left(m\cdot M_*k^{(1)}\right)}
{\left(k^{(1)}\cdot M_*k^{(1)}\right) \left(k^{(2)}\cdot M_*k^{(2)}\right)
-\left(k^{(1)}\cdot M_*k^{(2)}\right)^2}
J_F\\
J_{R_2,0}&=&\frac
{\left(k^{(1)}\cdot M_*k^{(2)}\right) \left(m\cdot M_*k^{(1)}\right)
-\left(k^{(1)}\cdot M_*k^{(1)}\right) \left(m\cdot M_*k^{(2)}\right)}
{\left(k^{(1)}\cdot M_*k^{(1)}\right) \left(k^{(2)}\cdot M_*k^{(2)}\right)
-\left(k^{(1)}\cdot M_*k^{(2)}\right)^2}
J_F~~.\nonumber
\end{eqnarray}
(the role of the elliptic structures formed around double resonances
in the Nekhoroshev theorem is discussed extensively in \cite{ben1999}).
If the higher order terms in the action variables of the development
of Eq.(\ref{nfdble0}) are taken into account, the constant energy
condition of Eq.(\ref{eneprime}) yields deformed ellipses on the plane
$(J_{R_1},J_{R_2})$. If $J_{R_1}\neq J_{R_{1,0}}$ or $J_{R_2}\neq J_{R_{2,0}}$,
the slow frequencies $\dot{\phi}_{R_1}\equiv\omega_{R_1}$,
$\dot{\phi}_{R_2}\equiv\omega_{R_2}$ are non-zero, and they
are given by
\begin{eqnarray}\label{omer12}
\omega_{R_1}&=&\left(k^{(1)}\cdot M_*k^{(1)}\right)(J_{R_1}-J_{R_{1,0}})
+\left(k^{(1)}\cdot M_*k^{(2)}\right)(J_{R_2}-J_{R_{2,0}})+ \ldots\\
\omega_{R_2}&=&\left(k^{(1)}\cdot M_*k^{(2)}\right)(J_{R_1}-J_{R_{1,0}})
+\left(k^{(2)}\cdot M_*k^{(2)}\right)(J_{R_2}-J_{R_{2,0}})+ \ldots\nonumber
\end{eqnarray}
On the other hand, due to the definition (\ref{restra}) one has
$$
\omega_{R_1}=k^{(1)}\cdot\omega(J^{(r_{opt})}),~~~
\omega_{R_2}=k^{(2)}\cdot\omega(J^{(r_{opt})}),~~~
\omega_{F}=m\cdot\omega(J^{(r_{opt})})
$$
which is valid for any value of $(J_{R_1},J_{R_2},J_F)$ in the
domain of convergence of the series (\ref{nfdble0}). It follows
that all the resonant manifolds defined by relations of the form
$(n_1k^{(1)}+n_2k^{(2)})\cdot\omega(J^{(r_{opt})})=0$ intersect
any of the planes $(J_{R_1},J_{R_2})$ corresponding to a fixed
value of $J_F$. Using the notation
$$
\Delta J_{R_i}=J_{R_i}-J_{R_{i,0}},~~~a_{ij}=k^{(i)}
\cdot M_*k^{(j)},~~~i,j=1,2
$$
the intersection of one resonant manifold with the plane
$(J_{R_1},J_{R_2})$ is a curve. In the linear approximation,
we have
\begin{eqnarray}\label{resmanjr}
 (n_1a_{11}+n_2a_{12})\Delta J_{R_1}
+(n_1a_{12}+n_2a_{22})\Delta J_{R_2}+ \ldots=0\nonumber
\end{eqnarray}
The above equation defines a `resonant line', which is the local
linear approximation to a `resonant curve'. All resonant lines
(or curves) pass through the point $(J_{R_{1,0}},J_{R_{2,0}})$,
which, therefore,  belongs to the resonant junction defined by
the wavevectors $k^{(1)},k^{(2)}$. To each resonant curve we can
associate a resonant strip in action space whose width is
proportional to the separatrix width for that resonance.
If, for a single pair of integers $(n_1,n_2)$, we only isolate
the resonant terms $g_{\pm n_1,\pm n_2}e^{\pm i(n_1\phi_{R_1}
+n_2\phi_{R_2})}$ in the normal form $Z$ (Eq.(\ref{hamres})),
we obtain a simplified resonant normal form $Z_{res(n_1,n_2)}$
corresponding to the limiting case of a single resonance.
In a strict sense, $Z_{res}$ describes well the dynamics far
from the resonant junction. However, it can also be used
in order to obtain estimates of the resonance width along the
whole resonant curve defined by the integer pair $(n_1,n_2)$.
To this end, the leading terms of $Z_{res(n_1,n_2)}$ are
(apart from constants):
\begin{eqnarray}\label{hamresn12}
Z_{res(n_1,n_2)} &=&\epsilon^{1/2}
\Bigg[{1\over 2}a_{11}\Delta J_{R_1}^2+a_{12}\Delta J_{R_1}\Delta J_{R_2}
+{1\over 2}a_{22}\Delta J_{R_2}^2+...\\
&+&\left(g_{n_1,n_2}e^{i(n_1\phi_{R_1}+n_2\phi_{R_2})}
+g_{-n_1,-n_2}e^{-i(n_1\phi_{R_1}+n_2\phi_{R_2})}\right)\Bigg]
+\ldots\nonumber
\end{eqnarray}
where the coefficients $g_{\pm n_1,\pm n_2}$ satisfy the estimate
\begin{equation}\label{gn12}
|g_{n_1,n_2}|\approx Ae^{-(|n_1||k^{(1)}|+|n_2||k^{(2)}|)\sigma}~~,
\end{equation}
due to Eq.(\ref{anal}). After still another transformation
$\Delta J_{R_1}=n_1J_R+n_2J_F$, $\Delta J_{R_2}=n_2J_R-n_1J_F$,
$\phi_R=n_1\phi_{R_1}+n_2\phi_{R_2}$,
$J_F$ becomes a second integral of motion of $Z_{res(n_1,n_2)}$,
which takes the form
\begin{eqnarray}\label{hamresn12n}
Z_{res(n_1,n_2)}&=&\epsilon^{1/2}\Bigg[c(J_F)
-{1\over 2}(a_{11}n_{11}^2+2a_{12}n_1n_2+a_{22}n_2^2)
(J_R-J_{R,0}(J_F))^2\nonumber\\
&+&\left(g_{n_1,n_2}e^{i\phi_{R}}
+g_{-n_1,-n_2}e^{-i\phi_{R}}\right)
+\ldots\Bigg]
\end{eqnarray}
where $c(J_F)$ and $J_{R,0}(J_F)$ are constants of the Hamiltonian
flow of (\ref{hamresn12n}). Combining (\ref{gn12}) and
(\ref{hamresn12n}), the separatrix width can be estimated as
\begin{equation}\label{sepwidth}
\Delta J_R \approx \sqrt{32Ae^{-(|n_1||k^{(1)}|+|n_2||k^{(2)}|)\sigma}
\over a_{11}n_{11}^2+2a_{12}n_1n_2+a_{22}n_2^2}~~.
\end{equation}
Eq.(\ref{sepwidth}) allows to estimate the width of a resonant strip
in the direction normal to a resonant curve on the plane
$(J_{R_1},J_{R_2})$. Using the relations $\Delta(\Delta J_{R_i})=
n_i\Delta J_{R}$ (for $\Delta J_F=0$), this estimate takes the form
\begin{equation}\label{reswidth}
\Delta J_{R,width}=
 \approx \left({32A(n_1^2+n_2^2)
\over a_{11}n_{11}^2+2a_{12}n_1n_2+a_{22}n_2^2}\right)^{1/2}
e^{-{1\over 2}(|n_1||k^{(1)}|+|n_2||k^{(2)}|)\sigma}~~.
\end{equation}

\begin{figure}
\centering
\includegraphics[scale=0.7]{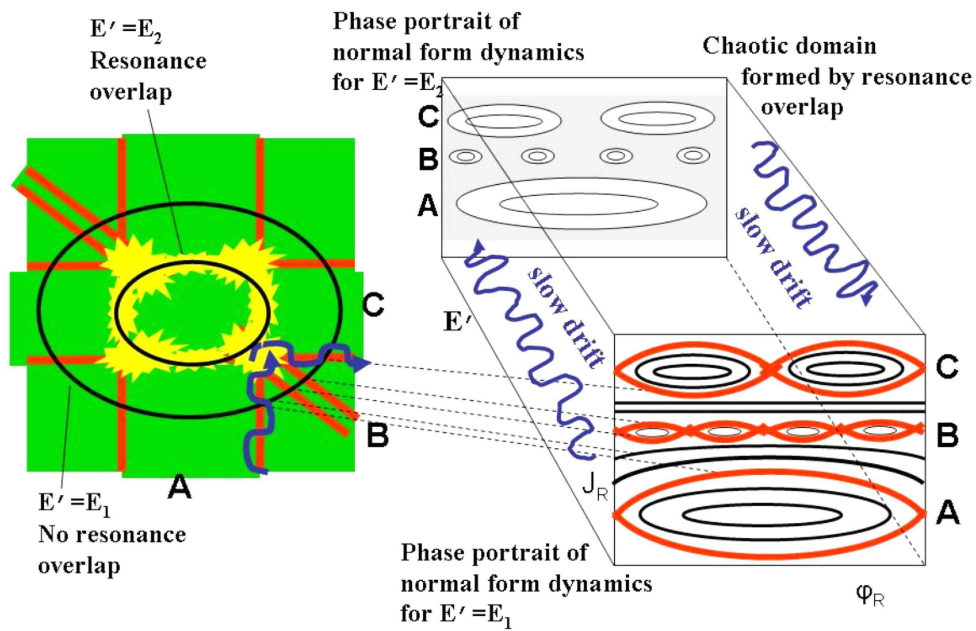}
\caption{Schematic representation of the normal form and remainder
dynamics in a domain of double resonance. Left panel: the resonant
structure formed in the action plane of the variables
$(J_{R_1},J_{R_2})$  by the overlapping of various
resonant strips whose limits (pairs of parallel red lines)
correspond to separatrix-like thin chaotic domains
around each resonance. Two constant normal form energy ellipses
$E'=E_1$ and $E'=E_2$ are also shown. Right: The front and back
panels show the phase portraits corresponding to a surface of
section (in one of the pairs $(\phi_{R_1},J_{R_1})$ or
$(\phi_{R_2},J_{R_2})$) under the normal form dynamics alone,
for the energies $E'=E_1$ (front panel) and $E'=E_2$ (back panel).
The blue curly arrows in both panels indicate the directions of
a possible `drift' motion (=slow change of the value of $E'$)
due to the influence of the remainder on dynamics.}
\label{fgdbleresmodel}
\end{figure}
The outcome of the analysis so far can be visualized with the help
of Figure \ref{fgdbleresmodel} (schematic). The left panel shows the
structure of a doubly-resonant domain in the plane of the resonant
action variables $(J_{R_1},J_{R_2})$. The two bold ellipses
correspond to the constant energy condition for two different values
of $E'$, namely $E'=E_1$ and $E'=E_2$ with $E_1>E_2$. Their common
center is the point $(J_{R_{1,0}}, J_{R_{2,0}})$  defined in
Eq.(\ref{jr0}). The three pairs of parallel red lines depict
the borders of the separatrix-like thin chaotic layers of three
resonances passing through the center. Infinitely many such
resonances exist, corresponding to different choices of integer
vectors $n\equiv(n_1,n_2)$; however, their width decreases as $|n|$
increases, according to Eq.(\ref{reswidth}). We thus show schematically
only three resonances with a relatively low value of $|n|$, named
by the letters `A', `B' and `C'. The blue curly curves indicate
a slow drift undergone by the chaotic orbits along the resonance
layers, allowing for a transition from one resonance to another.
This phenomenon, which will be addressed in detail below, is due
to the influence of the {\it remainder terms} of the normalized
Hamiltonian on dynamics. Here, however, we discuss first the (non-trivial)
influence of the {\it normal form terms} on dynamics, by considering
the Hamiltonian flow under the approximation $H\simeq Z$. Then, the
following facts hold: \\

\noindent
- For any fixed value of $E'$, and a fixed section in the angles,
the motion is confined on one ellipse. \\

\noindent - For $E'$ large enough ($E'=E_1$, outermost ellipse in
the left panel of Fig.\ref{fgdbleresmodel}), the various resonant
strips intersect the ellipse $E'=E_1$ at well distinct arcs, i.e.
there is no resonance overlap. The right front panel in
Fig.\ref{fgdbleresmodel} shows schematically the expected phase
portrait, which can be obtained by evaluating an appropriate
Poincar\'{e} surface of section, e.g. in the variables
$(J_{R_1},\phi_{R_1})$ or $(J_{R_2},\phi_{R_2})$. The dashed lines
show the correspondence between the limits of various resonant
domains depicted in the left and right panels. In particular, the
intersection of each resonant strip in the left panel with the
ellipse $E'=E_1$ corresponds to the appearance of an associated {\it
island chain} in the right panel. The size of islands is given
essentially by the separatrix width estimate of Eq.(\ref{reswidth}).
Hence, the size of the islands decreases exponentially with the
order of the resonance $n=|n_1|+|n_2|$. However, the main effect to
note is that, since all resonant strips are well separated on the
ellipse, the thin separatrix-like chaotic layers marking the borders
of each of their respective island chains do not overlap. As a
result the local chaos around one resonance is isolated from the
local chaos around the other resonances. In fact, the normal form
dynamics induces the presence of rotational KAM tori which, in this
approximation ($H\simeq Z$), completely obstruct the communication
among the resonances. Note that a detailed study of the dynamics of
the above type, induced by the doubly-resonant normal form, was
recently
presented in \cite{geletal2013}. \\

\noindent - Far from the domain of resonance overlap, the size of
the islands corresponding to each resonance is nearly independent of
the energy $E'$, as it depends essentially only on the size of the
Fourier coefficient of the corresponding harmonics in the
Hamiltonian. However, the separation of the islands is reduced as
the energy {\it decreases}, since this separation is given
essentially by the separation between the distinct arcs in
Fig.\ref{fgdbleresmodel} at which the various resonances intersect
the ellipse corresponding to a fixed energy $E'$. As a result, below
a critical energy $E'_c$, significant resonance overlap takes place,
leading to the communication of the chaotic layers of the various
resonances and an overall increase of chaos. This is shown in the
left panel of Fig.\ref{fgdbleresmodel} for an ellipse $E'=E_2<E_c'$,
with the corresponding phase portrait shown in the right back panel.
We note in particular the `merging' of all three resonant domains
one into the other, which produces a large connected chaotic domain
surrounding all three island chains (and many other smaller chains,
not visible in this scale).

The value of the critical energy $E_c'$ marking the onset of large
scale resonance overlap can be estimated as follows: Each resonant
strip intersects one fixed energy ellipse on one arc segment.
Also, Eq.(\ref{reswidth}) can be replaced by the estimate
\begin{equation}\label{reswdk}
\Delta J_{R,width} \approx {(32A)^{1/2}\over M_hk_{1,2}}
e^{-{1\over 2}nk_{1,2}\sigma}
\end{equation}
where $n=|n_1|+|n_2|$, $k_{1,2}=(|k^{(1)}|+|k^{(2)}|)/2$, and
$M_h=(\mu_{min}+\mu_{max})/2$, with the constants $\mu_{min},\mu_{max}$
defined as in Eq.(\ref{mh}). The total length $S_{res}$ of all
segments can be now estimated by summing, for all $n$,
the estimate (\ref{reswdk}), namely
\begin{equation}\label{sres}
S_{res}\approx
{(32A)^{1/2}\over M_hk_{1,2}}
\sum_{n=1}^{\infty}e^{-{1\over 2}nk_{1,2}\sigma}
\approx
{(128A)^{1/2}\over M_hk_{1,2}\sigma}e^{-{1\over 2}k_{1,2}\sigma}
\end{equation}
On the other hand, the total circumference of the ellipse for the energy $E'$
is estimated as $S_{E'}=\pi R(E')^2$ where $R(E')$ is the geometric mean of the
ellipse's major and minor semi-axes. For $R(E')$ one has the obvious
estimate $R(E')\sim (2E'/(\epsilon^{1/2}M_h))^{1/2}$, whence
\begin{equation}\label{sep}
S_{E'}\sim {2\pi E'\over \epsilon^{1/2}M_h}~~~.
\end{equation}
The critical energy $E'=E_c'$ can now be estimated as the value where
$S(E')\approx S_{res}$, implying that the associated ellipse is fully
covered by segments of resonant strips. Thus
\begin{equation}\label{epcrit}
E_c'\approx {32(\epsilon A)^{1/2}
\over \pi k_{1,2}\sigma}e^{-{1\over 2}k_{1,2}\sigma}~~.
\end{equation}
Eq.(\ref{epcrit}) implies that $E_c'$ is a
$O(\epsilon^{1/2}e^{-{1\over 2}k_{1,2}\sigma})$ quantity.

So far, we have neglected the role of the remainder in dynamics.
In Fig.\ref{fgdbleresmodel}, the drift in action space caused by the
remainder is shown schematically by the blue curly curves in both
the left and right panels. Their significance is the following:
The energy $E=h$ corresponding to the total Hamiltonian
$h=Z+R^{(r_{opt})}$ of Eq.(\ref{hamres2}) is an exactly preserved
quantity. Thus, the doubly-resonant normal form energy $E'$ as well
as $J_F$ cannot be preserved exactly, but they are approximate integrals,
i.e. they undergo time variations bounded by an $O(||R^{(r_{opt})}||)$
quantity. In Fig.\ref{fgdbleresmodel}, such variations will
in general lead to a very slow change of the value of $E'$,
i.e. a very slow drift of the chaotic orbits from one ellipse
to another. We seek to estimate the time required for the
remainder to induce a transition between two ellipses with
an energy difference of the same order as $E_c'$, namely
\begin{equation}\label{enedif}
E_2'-E_1'=O(\epsilon^{1/2}e^{-{1\over 2}k_{1,2}\sigma})
\end{equation}
assuming that this effect can be described as a {\it random walk}
in the value of $E'$ (numerical evidence for this assumption
will be provided in section 3). Let $T$ be an average period
of the oscillations of the resonant variables. By
Eqs.(\ref{hamresn12}) and (\ref{gn12}), the estimate
$T\sim(\epsilon A)^{-1/2}e^{n_{eff}k_{1,2}\sigma/2}$
holds, for a constant $n_{eff}\sim 1$ marking the order of
the most important resonances in (\ref{hamresn12}).
In consecutive steps, $d E'$ can be either positive or negative,
while its typical size is $|d E'|\sim ||R_{opt}||$. Then, after
$N$ steps of a random walk (in the values of $E'$), we find
an rms spread of these values given by
\begin{equation}\label{derms}
\Delta E\approx N^{1/2} ||R_{opt}||
\end{equation}
Using (\ref{enedif}) and (\ref{derms}), the number of steps required
for the spread $\Delta E$ to become equal to $E_2'-E_1'$ (given
by (\ref{enedif}) is $N\sim \epsilon e^{-n_{eff}k_{1,2}\sigma}
||R_{opt}||^{-2}$. The diffusion coefficient can be estimated as
\begin{equation}\label{difcfest}
D\sim {\Delta E^2\over NT}\sim
\left(\epsilon A e^{-n_{eff}k_{1,2}\sigma}\right)^{1/2}
||R_{opt}||^2
\end{equation}
i.e. the diffusion coefficient scales as the square of the size
of the optimal remainder function. This relation is probed by
detailed numerical experiments in section 3.

\subsubsection{Simple resonance}
When $k^{(1)}<K_{opt}(\epsilon)\leq k^{(2)}$, $I_*$ is
simply-resonant with respect to the optimal K--truncation.
In this case, the normal form contains terms either independent
of the angles, or depending on them via trigonometric terms
of the form $\exp(ink^{(1)}\cdot\phi^{(r_{opt})})$,
$n\in {\cal Z}^*$. Using the same notations as in the previous
subsection, the transformed Hamiltonian reads:
\begin{eqnarray}\label{hamressple0}
h(J^{(r_{opt})},\phi^{(r_{opt})})&=&Z(J^{(r_{opt})},\phi^{(r_{opt})})
+R(J^{(r_{opt})},\phi^{(r_{opt})})\nonumber\\
&=&\omega_*\cdot J^{(r_{opt})}
+ \epsilon^{1/2}\sum_{i=1}^3\sum_{j=1}^3
{1\over 2}M_{ij*}J^{(r_{opt})}_iJ^{(r_{opt})}_j +\ldots\\
&+&\epsilon^{1/2}\sum_{n\in {\cal Z}^*}g_n(J^{(r_{opt})})
\exp(i(nk^{(1)}\cdot\phi^{(r_{opt})})+\ldots\nonumber\\
&+&R(J^{(r_{opt})},\phi^{(r_{opt})})\nonumber
\end{eqnarray}
\begin{figure}
\centering
\includegraphics[scale=0.5]{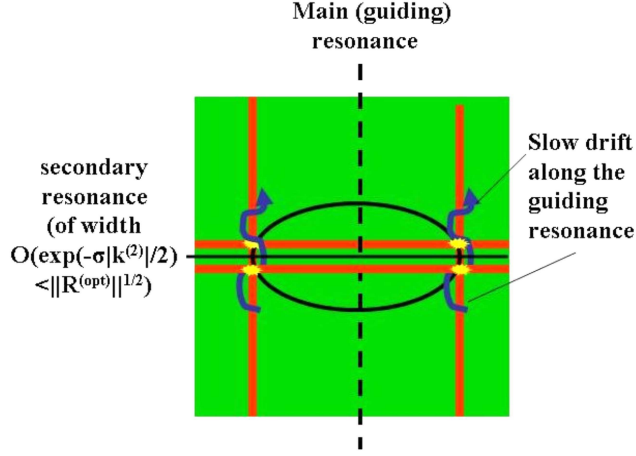}
\caption{Same as in the left panel of Fig.\ref{fgdbleresmodel},
but for a simple resonance. In this case, any other resonance
crossing the main (guiding) resonance has an exponentially small
width and acts as a `driving' resonance for diffusion.}
\label{fgspleresmodel}
\end{figure}
Repeating all steps as in the case of double resonance leads to the
normal form
\begin{eqnarray}\label{hamresn1}
Z_{res} &=& {1\over 2}a_{11}\Delta J_{R_1}^2+a_{12}\Delta J_{R_1}\Delta J_{R_2}
+{1\over 2}a_{22}\Delta J_{R_2}^2+...\\
&+&\epsilon^{1/2}\left(g_ne^{in\phi_{R_1}}
+g_{-n}e^{-in\phi_{R_1}}\right)
+\ldots\nonumber
\end{eqnarray}
The main difference with respect to the doubly-resonant normal form
(\ref{hamresn12}) is that, the angle $\phi_{R_2}$ being ignorable, the
action $J_{R_2}$ (or $\Delta J_{R_2}$) is an integral of the flow of
$Z_{res}$, in addition to $J_F$. Thus, $Z_{res}$ defines an integrable
Hamiltonian. A pair of constant values $J_F=c_1$, $\Delta J_{R_2}=c_2$
defines a straight line
\begin{equation}\label{resline}
\Delta J_{R_1} = -{a_{12}\over a_{11}}c_2
\end{equation}
which corresponds to the unique resonance
$\omega_{R_1}(J^{(r_{opt})})=k^{(1)}\cdot\omega(J^{(r_{opt})})=0$.
This will be called `main resonance' (= the `guiding resonance'
in \cite{chi1979}). In Figure \ref{fgspleresmodel} (schematic),
the domain of the main resonance is delimited by two vertical
thick red lines corresponding to the separatrix-like thin chaotic
layers at the boundary of the resonance similarly to
Fig.\ref{fgdbleresmodel}. Using similar arguments as in the
derivation of Eq.(\ref{reswdk}), the separatrix width
can be estimated as
\begin{equation}\label{reswdk1}
\Delta J_{R,width} \approx {(32A)^{1/2}\over M_h|k^{(1)}|}
e^{-{1\over 2}|k^{(1)}|\sigma}~~~.
\end{equation}
Under the normal form dynamics, motions are allowed only across the
resonance, i.e. in the direction $\Delta J_{R_2}=const$.
In Fig.\ref{fgspleresmodel} this is the horizontal direction.
The thin strip delimited by two horizontal red lines corresponds
to the resonance with resonant wavevector $k^{(2)}$, which,
since $k^{(2)}>K(\epsilon)$, is now of width exponentially
small ($O(\epsilon^{1/2}e^{-\sigma|k^{(2)}|/2}$). Thus, it
will be called a `secondary' resonance.

In order to estimate the speed of diffusion as a function of
the optimal remainder in this case, let us note first that the
influence of the remainder on dynamics is to slowly change the
value of the two approximate integrals $J_F$ and $\Delta J_{R_2}$,
that would be exactly preserved under the normal form dynamics.
In view of Eq.(\ref{hamresn1}), the Hamiltonian (\ref{hamressple0})
can be approximated by
\begin{eqnarray}\label{hamressple}
h&\approx&(m\cdot\omega_*)J_F
+\epsilon^{1/2}\Bigg[{1\over 2}a_{11}\Delta J_{R_1}^2
+a_{12}\Delta J_{R_1}\Delta J_{R_2}
+{1\over 2}a_{22}\Delta J_{R_2}^2+...\nonumber
+2f_{R_1}\cos(\phi_{R_1})+\ldots\\
&+&\sum_{|k|\geq K^{(opt)}}f_{k*}
\exp[ik\cdot(\kappa_1\phi_{R_1}+\kappa_2\phi_{R_2}+
\kappa_3\phi_{F})]+...\Bigg]
\end{eqnarray}
where i) the (non-integer) vectors $\kappa_i$, $i=1,2,3$ come from
the solution of the right Eqs.(\ref{restra}) for the angles
$\phi_i^{(r_{opt})}$ in terms of the angles $\phi_{R_1}$, $\phi_{R_2}$,
and $\phi_{F}$, and ii) we approximate all the Fourier coefficients
in the remainder series by their constant values $f_{k*}$ at the
points $\Delta J_{R_1}=\Delta J_{R_2}=0$ (we set $f_{R_1}=f_{k*}$
for $k=k^{(1)}$).

The latter approximation is sufficient for estimates regarding the
speed of diffusion. The key remark is that for all the coefficients
$f_{k*}$ the bound $|f_{k*}|<||R_{opt}||$ holds, while, for the
leading Fourier term $\exp(ik_l\cdot\phi^{(r_{opt})})$ in the
remainder we have $|f_{k_l*}|\sim||R_{opt}||$. In fact, we typically
find that the size of the leading term is larger from the size of
the remaining terms by several orders of magnitude, since this term
contains a repeated product of small divisors of the form
$k_l\cdot\omega_*$ (see Appendix A). Furthermore, using an analysis
as in \cite{eftetal2004}, we readily find $|k_l|=(1-d)K_{opt}$,
where $0<d<1$ is a so-called (in \cite{eftetal2004}) `delay'
constant. We note in passing that the Fourier terms of the form
$\exp(ik_l\cdot\phi^{(r_{opt})})$ are called `resonant' in
\cite{morgio1997}. The value of the diffusion coefficient can now be
estimated by applying the  heuristic theory of Chirikov
(\cite{chi1979}, see also \cite{cin2002} and \cite{cachetal2010}) in
the Hamiltonian model (\ref{hamressple}). The estimate
\begin{equation}\label{difchi}
D\sim {\epsilon\over 2\Omega_G^2 T}|f_{k_l*}|^2A(|\kappa_l|)^2
\end{equation}
holds, where $\Omega_G=\epsilon^{1/4}f_{R_1}^{1/2}$,
$T=ln(32e/w)/\Omega_G$ is an average period of motion within
the main resonance separatrix-like thin chaotic layer, of
width $w$, $A$ is the Melnikov function with argument
$|\kappa_l|$ (see Appendix B of \cite{fermel2007}),
the vector $\kappa_l$ being defined by the relation
$\kappa_{l,1}\phi_{R_1}+\kappa_{l,2}\phi_{R_2}+
\kappa_{l,3}\phi_F = k_l\cdot\phi^{(r_{opt})}$.
The estimate $A(|\kappa_l|)\sim 8\pi|\kappa_l| e^{-\pi|\kappa_l|/2}$
holds. In view of Eq.(\ref{restra}) however, we have that
$|\kappa_l|=O((1-d)K^{(opt)}/|k^{(1)}|)$. Since $K_{opt}\sim
\epsilon^{-1/4}$ (see Appendix B), and $||R_{opt}||\sim e^{-\sigma K^{(opt)}}$,
it follows that $A(|\kappa_l|)\sim \epsilon^{3/4}||R_{opt}||^b$,
for an exponent $b>0$. Putting these estimates together, we
finally arrive at a steeper dependence of the diffusion coefficient
$D$ on the optimal remainder $||R_{opt}||$ in the case of
simple resonance than in the case of double resonance, namely:
\begin{equation}\label{difcfsple}
D\sim {\epsilon\over 2\Omega_G^2 T}\epsilon^{3/4}||R_{opt}||^{2(1+b)}
\end{equation}

Regarding now the precise value of $b$, it is hardly
tractable to determine this on the basis exclusively of the
behavior of the Melnikov integrals discussed above. We note,
however, that the quantity $A(\kappa_l)$ yields the size of
the `splitting'$S$ of the separatrix of the main (guiding)
resonance due to the effects of the leading term in the
remainder function. The relation between the separatrix
splitting and the size of the optimal remainder has been
examined in \cite{nei1984} and later in \cite{morgio1997}.
In the latter work, the estimate $S\sim\mu^{1/2}$ was
predicted and probed by numerical experiments, where
$\mu$ (in the notation of \cite{morgio1997}) is the
effective size of the perturbation to the normal form
pendulum dynamics caused by the remainder. Setting thus
$\mu\sim||R_{opt}||$ suggests the scaling $A(\kappa_l)\sim
S\sim ||R_{opt}||^{1/2}$, whereby the constant $b$ can be
estimated as $b\simeq 1/2$. Hence (in view of \ref{difcfsple})
$$
D\sim ||R_{opt}||^3
$$
in simply resonant domains.

Despite the heuristic character of the above derivation, it
seems that the value $b\simeq 1/2$ is supported by the results
of numerical experiments. In particular, in \cite{eft2008} the
diffusion coefficient $D$ along a simple resonance was compared
directly to the size of the optimal normal form remainder. It was
found that $D\propto ||R_{opt}||^{2.98}$, essentially confirming that
$p=2(1+b)\simeq 3$. We point out, however, that in \cite{legetal2010a}
a different exponent was found $p\simeq 2.56$ regarding the same
resonance as in \cite{eft2008}, while it was found that $p=2.1$
in the case of a very low order simple resonance (with $|k^{(1)}|<K'$),
which is not discussed in our present work. These exponents, on
the other hand, depend on the chosen definition of the numerical
measure used to estimate both $S$ and $||R_{opt}||$. Thus, a
detailed quantitative comparison of the works cited above is
left as on open problem for future study.

\section{Numerical results}
In our numerical work we employ the same Hamiltonian model of three
degrees of freedom as in \cite{froetal2000,guzetal2002,legetal2003,
guzetal2005}. The Hamiltonian reads:
\begin{equation}\label{hamfr}
H = H_0+\epsilon H_1 = {I_1^2+I_2^2\over 2}+I_3 +
{\epsilon\over 4+\cos\phi_1+\cos\phi_2+\cos\phi_3}~~~.
\end{equation}
This model has a particularly simple, yet sufficient for our purpose,
structure, allowing to probe numerically all steps of the previous
section. In particular:

\subsection{Analyticity and convexity}
The function (\ref{hamfr}) is polynomial in the action variables,
thus it is analytic in any complex extension of ${\cal I}={\cal R}^3$.
On the other hand, the domain of analyticity in the angle variables was
examined in \cite{eft2008}. It was found that analyticity can be
established in a set $Re(\phi_i)\in{\cal T}$, $Im(\phi_i)<\sigma$, $i=1,2,3$,
for a positive constant $\sigma$ estimated semi-analytically as $\sigma\simeq
0.82$. Accordingly, the coefficients $h_k$ of the Fourier development
\begin{equation}\label{fourier}
{1\over 4+\cos\phi_1+\cos\phi_2+\cos\phi_3} =
\sum_{k_1=-\infty}^{\infty}
\sum_{k_2=-\infty}^{\infty}
\sum_{k_3=-\infty}^{\infty}
h_k\exp(ik\cdot\phi)
\end{equation}
where $k\equiv(k_1,k_2,k_3)$, $\phi\equiv(\phi_1,\phi_2,\phi_3)$,
decay exponentially. The distance of the nearest singularity, with
respect to each of the angles $\phi_i$, from the real axis is given
by the solution of $cos\phi=-4/3$, or $\phi=\pi+0.795365i$. Thus,
the following bound holds:
\begin{equation}\label{expdecay}
|h_k|\leq A\exp(-|k|\sigma),~~~A\simeq 0.05,~~\sigma=0.795365~~.
\end{equation}

As regards convexity, for all $I_*\in{\cal I}$ the matrix $M_*$ has a
particularly simple structure, since we have $M_{11*}=M_{22*}=1$, and $M_{ij*}=0$
for all other $i,j$. Thus there are two positive eigenvalues equal to unity
and one equal to zero, while $\mu_{min}=\mu_{max}=1$.

The constant energy condition $E=(I_1^2+I_2^2)/2+I_3$ defines a paraboloid
in the action space. The resonant manifolds are planes, since $\omega_1=I_1$,
$\omega_2=I_2$, $\omega_3=1$, whereby the resonance conditions
\begin{equation}\label{frfr}
k_1\omega_1+k_2\omega_2+k_3\omega_3=k_1I_1+k_2I_2+k_3=0
\end{equation}
for all $k\equiv(k_1,k_2,k_3)$ define planes normal to the $(I_1,I_2)$
plane. It follows that, when projected to the $(I_1,I_2)$ plane,
the intersections of all resonant manifolds with a
surface of constant energy of the unperturbed problem yield a set of
straight lines. This greatly facilitates the numerical study, since
all diffusing orbits in the perturbed problem follow piecewise straight
paths nearly parallel to one or more resonant lines of the unperturbed
problem, while the orbits can only change direction by approaching close
to resonance junctions. Examples of diffusion of this type along a simple
resonance where studied in \cite{legetal2003}, while the case of consecutive
encounters with doubly-resonant domains was examined in a mapping model
\cite{guzetal2005} variant of the Hamiltonian model (\ref{hamfr}).

\subsection{Normal form construction and optimal remainder}
The connection between the size of the optimal remainder $||R_{opt}||$
and the diffusion coefficient $D$ in a case of simple resonance was
the main subject of a previous study \cite{eft2008}. Following
the same terminology and notations as in section 2 above,
the point $I_*$ in the normal form construction in \cite{eft2008}
was chosen as $(I_{1*},I_{2*},I_{3*})=(0.31,0.155,1)$. For this point
we have (viz. Eq.(\ref{mvec})) $k^{(1)}=(1,-2,0)$, $k^{(2)}=(100,0,-31)$,
$m=(31,155,100)$. The optimal truncation order in all calculations of
\cite{eft2008} varied from $K_{opt}(\epsilon)=18$ to $K_{opt}(\epsilon)=39$
(depending on the value of $\epsilon$ in the range considered).
Thus, in all cases we have $|k^{(1)}|<K_{opt}(\epsilon)<|k^{(2)}|$,
that is the so-chosen point $I_*$ was found to be simply resonant with respect
to the optimal K--truncation. Following Fig.5 of \cite{eft2008} it was then
found by numerical fitting that the diffusion coefficient $D$ scales with
the optimal remainder as $D\propto ||R_{opt}||^{2.98}$. A theoretical
justification for this `steepening' of the power-law with respect to
the exponent $p\simeq 2$ holding in double resonances was given in
subsection 2.3.2.

In order to probe now the dependence of $D$ on $||R_{opt}||$ in the
case of a double resonance, in the sequel we focus our numerical study
on a different point of ${\cal D}$, namely $(I_{1*},I_{2*},I_{3*})=(0.4,0.2,1)$.
The basic resonant wavevectors are
\begin{equation}\label{resvectors}
k^{(1)}=(1,-2,0),~~~k^{(2)}=(2,1,-1),~~~\mbox{implying}~~m=(2,1,5)~.
\end{equation}
The Hamiltonian normalization is carried out as exposed in subsection 2.2.
The interval of values of $\epsilon$ considered is $0.001\leq\epsilon\leq 0.02$
which, according to \cite{legetal2003} is below the critical value for the
onset of the `Nekhoroshev regime' ($\epsilon_c\simeq 0.03$). Furthermore, it
will be shown below that for all values of $\epsilon$ in the above interval
the optimal Fourier-truncation order $K_{opt}$ turns to be much larger than
$K=4$. On the other hand, for the basic wavevectors we have $|k^{(1)}|=3$,
$|k^{(2)}|=4$. Thus, for all considered values of $\epsilon$ one has
$|k^{(1)}|<|k^{(2)}|<K_{opt}(\epsilon)$, that is, the point $I_*$ is
doubly resonant with respect to any of the optimal K--truncations
considered in the sequel.

Due to Eq.(\ref{kprime}), the constant $K'$ in terms of which
book-keeping is implemented changes with $\epsilon$. However, one notices
that, because of the logarithmic dependence of $K'$ on $\epsilon$, in the
largest part of the interval $0.001\leq\epsilon\leq 0.02$, where we focus,
one has a constant value $K'=3$, while one has $K'=2$ only close to
the upper limit $\epsilon=0.02$ and $K'=4$ close to the lower limit
$\epsilon=0.001$. For simplicity, we thus fixed the value of $K'$ as $K'=3$
in all normal form computations. Doing so, computer memory limitations
restrict all computed expansions to a maximum order $r_{max}=17$ in the
book-keeping parameter $\lambda$, or maximum order $|k|_{max}=17K'-1=50$
in Fourier space. In fact, for $\epsilon> 0.005$ we perform at most 14
normalization steps, so that the remainder contains terms of at least three
consecutive orders in $\lambda$, namely $r=15$, $16$ and $17$. As explained
below, this allows us to perform some numerical tests regarding the
convergence of the remainder series when the optimal normalization order
is as high as $r_{opt}$=14 (or $K_{opt}=42$). On the other hand, for
$\epsilon\leq 0.005$ we allow for one more normalization step ($r=15$)
in order to get as close as possible to the optimal order, which, as shown
below for $\epsilon<0.004$ is larger than 14. Thus, for the calculation of
the corresponding remainder value at this order ($r=15$) we necessarily
have to rely on the sum of only two rather than three or more consecutive
terms.

Writing the truncated (at order 17) remainder function as:
\begin{equation}\label{remrtr}
R^{(r)}(J^{(r)},\phi^{(r)})_{\leq 17}=
\sum_{s=r+1}^{17} R^{(r)}_s(J^{(r)},\phi^{(r)})~,
\end{equation}
where $R^{(r)}_s(J^{(r)},\phi^{(r)})$ are the terms of order $s$ in the
book-keeping parameter $\lambda$ allows us to probe numerically the
convergence of the remainder function within any chosen domain
${\cal W}_{I_*,B}$ in action space. To this end, at any normalization
order $r$, let us consider a disk $(J_1^{(r)})^2+(J_2^{(r)})^2\leq \rho^2$
in the space of the transformed action variables
(we neglect the action $I_3$ which, in the particular case of the Hamiltonian
(\ref{hamfr}), is dummy, i.e. it does not appear in any higher order term
of either the normal form or the remainder). This is a deformed disk also
in the old canonical variables $J_1,J_2$, limited by a boundary given
approximately by
$\epsilon(J_1^2+J_2^2)=(I_1-I_{1*})^2+(I_2-I_{2*})^2\simeq \epsilon \rho^2$
(cf. the action re-scaling given by Eq.(\ref{resc})).
For the Hamiltonian (\ref{hamfr}) one can readily check that all the terms
in $R^{(r)}_s(J^{(r)},\phi^{(r)})$ are trigonometric polynomials of maximum
degree $K's-1=3s-1$ whose coefficients are polynomial of maximum degree $s-1$
in the actions, namely
\begin{equation}\label{remsexp}
R^{(r)}_s(J^{(r)},\phi^{(r)})=\sum_{|k|=0}^{3s-1}
\left(\sum_{q=0}^{s-1}\sum_{l=0}^q
R_{k,l,q-l}(J_1^{(r)})^l(J_2^{(r)})^{q-l}\right)\exp(ik\cdot\phi^{(r)})~~~.
\end{equation}
Furthermore, in the disk ${\cal W}_{I_*,\epsilon^{1/2}\rho}$ the obvious bound
\begin{equation}\label{jball}
\sup_{{\cal W}_{I_*,\epsilon^{1/2}\rho}}|(J_1^{(r)})^l(J_2^{(r)})^{q-l}|=
\left({l^{l/2}(q-l)^{(q-l)/2} \over q^{q/2}}\right)\rho^q
\end{equation}
holds. We thus define the norm
\begin{equation}\label{remsnorm}
||R^{(r)}_s(J^{(r)},\phi^{(r)})||_{{\cal W}_{I_*,\epsilon^{1/2}\rho}}
=\sum_{|k|=0}^{3s-1}\sum_{q=0}^{s-1}\sum_{l=0}^q
|R_{k,l,q-l}|\left({l^{l/2}(q-l)^{(q-l)/2} \over q^{q/2}}\right)\rho^q~~
\end{equation}
in view of which a numerical estimate of the size of the remainder
within ${\cal W}_{I_*,\rho}$ can be obtained. In fact, by calculating
the truncated sums
\begin{equation}\label{sumrem}
||R^{(r)}(J^{(r)},\phi^{(r)})||_{\leq p,{\cal W}_{I_*,\epsilon^{1/2}\rho}}
=\sum_{s=r+1}^{p} ||R^{(r)}_s(J^{(r)},\phi^{(r)})||_{{\cal W}_{I_*,\epsilon^{1/2}\rho}}
\end{equation}
for any fixed choice of $\rho$, where $p$ takes all values
$p=r+1,r+2,\ldots,17$, we can have a clear numerical indication
of whether the remainder function was calculated up to a sufficiently
high order for convergence to have been practically reached.

\begin{figure}
\centering
\includegraphics[scale=0.5]{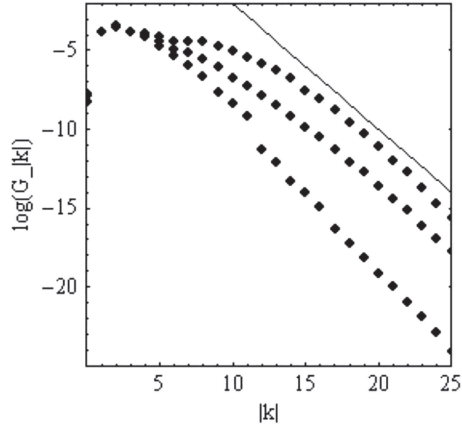}
\caption{The logarithm of the quantity $G_{|k|}^{(r)}$ (see text)
as a function of the Fourier order $|k|$, for $\epsilon=0.01$,
at the normalization orders $r=4$, $r=8$, and $r=11$
(lower, middle and upper set of points respectively).
All three curves exhibit an exponential decay for
large $|k|$, with nearly the same asymptotic law.
The straight line has inclination $\sigma=-0.8$.}
\label{fggksigma}
\end{figure}

The maximum value of $\rho$ for which the series $||R^{(r)}(J^{(r)}
,\phi^{(r)})||_{\leq\infty,{\cal W}_{I_*,\epsilon^{1/2}\rho}}$
converges absolutely sets the size of the doubly-resonant domain
$B=\epsilon^{1/2}\rho_{max}$ (in non-scaled variables) where the
normal form calculations are valid. In practice, we are interested
in the diffusion of orbits with initial conditions inside this
domain. In particular, in subsection 3.2 we will consider orbits
starting on the circle $\rho_0=0.27$.  All our numerical orbits are
studied up to a time in which their distance from the center of the
double resonance changes significantly less than
$\Delta\rho=10^{-1}$ (see below). Variations of this order at
maximum are found when we measure $\rho$ either in the original
canonical action variables or in the variables after the optimal
canonical transformation. Thus, for all the orbits we can set a safe
outer boundary $\rho<\rho_b=0.4$ within which they are well
confined. We then verify numerically that this domain belongs to the
analyticity domain of the various transformations employed in the
form of series (of the new variables in terms of the old variables
or vice versa). This check is made by finding whether the Fourier
coefficients of the series exhibit an exponential decay. An example
is given in Fig.\ref{fggksigma}. We consider the Fourier series
yielding the new transformed canonical action $J_1^{(r)}$ as a
function of the old canonical variables, for $\epsilon=0.01$, at the
normalization orders $r=4,8$ and $11$. Writing this as a series
\begin{equation}\label{gkser}
J_1^{(r)}=J_1+\sum_k \sum_{s_1,s_2=0}^{s_{max}(k)}
g_{k,s_1,s_2}^{(r)} J_1^{s_1}J_2^{s_2}\exp(ik\cdot\phi)
\end{equation}
we define the coefficients
\begin{equation}\label{gkcoef}
G_{|k|}^{(r)}=\sum_{k_1,k_2,~|k_1|+|k_2|=|k|}
\sum_{s_1,s_2=0}^{s_{max}(k)}
|g_{k,s_1,s_2}|{s_1^{s_1/2}s_2^{s_2/2}
\over (s_1+s_2)^{(s_1+s_2)/2}}
\rho_b^{s_1+s_2}
\end{equation}
Figure \ref{fggksigma} shows the coefficients $G_|k|^{(r)}$ for
$\epsilon=0.01$, $\rho_b=0.4$, and $r=4,8$ and $11$. We observe that
all three curves exhibit a tail showing exponential decay of the
Fourier coefficients. However, it is remarkable that the asymptotic
exponential slope seems to change only marginally. Instead, the main
change, as $r$ increases, regards that formation of a `plateau' of
Fourier coefficients of nearly constant size formed for small $|k|$.
Namely, the width of the plateau increases as $r$ increases. It is
remarkable that the asymptotic tail laws for all $r$ appear to
follow an exponential decay with the same constant $\sigma\simeq
0.8$, i.e. with nearly the same value as the constant appearing in
the analyticity condition of the original Hamiltonian (cf.
Eq.(\ref{expdecay})). This effect shows that, while in the usual
proofs of the Nekhoroshev theorem one requires a reduction of the
analyticity domain at every normalization step, i.e. one considers
bounds of the form $G_|k|^{(r)} \leq A^{(r)}e^{-\sigma_r|k|}$ with
$\sigma_r<\sigma_{r-1} <\ldots<\sigma_1$, in practice the dependence
of the coefficients $G_{|k|}^{(r)}$ on $|k|$ is more complicated
than a simple exponential decay law. In fact, the constants
$\sigma_r$ reflect an average exponential slope that compensates
between the plateau, for small $|k|$, and the exponential tail, for
large $|k|$. Namely, as the width of the plateau increases with $r$,
one obtains smaller and smaller values of the average exponential
decay constant $\sigma_r$.

\begin{figure}
\centering
\includegraphics[scale=0.7]{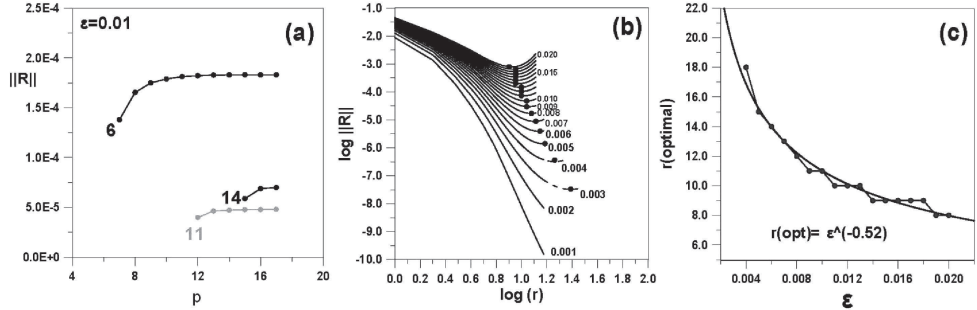}
\caption{(a) The value of the remainder norm
$||R^{(r)}||_{\leq p,{\cal W}_{I_*,\epsilon^{1/2}\rho_0}}$
as a function of the truncation order $p$ when $\epsilon=0.01$,
$\rho_0=0.27$, and the normalization orders are $r=6$ (upper curve),
$r=11$ (lower curve) and $r=14$ (middle curve). (b) The value of
$||R^{(r)}||_{\leq 17,{\cal W}_{I_*,\rho_0}}$ as a function of $r$
for different values of $\epsilon$. For $\epsilon=0.004$ and
$\epsilon=0.003$, the dashed curves after the order $r=15$ are
found by quadratic extrapolation. No attempt to extrapolate was
made for $\epsilon=0.002$ and $\epsilon=0.001$. (c) The optimal
normalization order $r_{opt}$ as a function of $\epsilon$ together
with a power-law best fitting curve.}
\label{fgremopt}
\end{figure}
Fig.\ref{fgremopt}a shows now an
example of the behavior of the truncated remainder function for
$\rho=\rho_0$ and $\epsilon=0.01$. The upper curve shows the value
of $||R^{(r)}(J^{(r)}
,\phi^{(r)})||_{\leq p,{\cal W}_{I_*,\epsilon^{1/2}\rho_0}}$
at the normalization order $r=6$ as a function of $p$ for $p=7,...17$.
Clearly, after $p=9$ the cumulative sum (\ref{sumrem}) shows no
further substantial variation, which indicates that the remainder
series converges after three consecutive terms $p=7,8$ and $9$
(this is verified also by computing numerically a convergence
criterion like d'Alembert's criterion). The lower and middle curves
show now the same effect for the normalization orders $r=11$ and
$r=14$ respectively. Note that the three consecutive truncation
orders $p=15,16$ and $17$ allowed for the computation of the
remainder at the normalization order $r=14$ are essentially
sufficient to demonstrate the convergence of the remainder.
Hence, $||R^{(r)}(J^{(r)},
\phi^{(r)})||_{\leq 17,{\cal W}_{I_*,\epsilon^{1/2}\rho_0}}$
represents a good numerical estimator of the value of the
remainder series for normalization orders up to $r=14$. However,
the main effect to note is that the estimated remainder value
$||R^{(r)}(J^{(r)}
,\phi^{(r)})||_{\leq 17,{\cal W}_{I_*,\epsilon^{1/2}\rho_0}}$
found for $r=14$ is larger than the one for $r=11$, implying that
the {\it optimal} normalization order $r_{opt}$ is below $r=14$.
Fig.{\ref{fgremopt}}b shows, precisely, the asymptotic character
of the above normalization, showing $||R^{(r)}(J^{(r)}
,\phi^{(r)})||_{\leq 17,{\cal W}_{I_*,\epsilon^{1/2}\rho_0}}$ against
the normalization order $r$ for various values of $\epsilon$ as indicated
in the figure. For all values down to $\epsilon=0.005$ we now observe
the asymptotic behavior, namely the size of the remainder initially
decreases as $r$ increases, giving the impression that the
normalization might be a convergent procedure. However, this
trend is reversed after an optimal order $r_{opt}$, where the
remainder reaches its minimum value, while, for $r>r_{opt}$
the remainder increases with $r$ and eventually goes to infinity.
We also observe that for $\epsilon\leq 0.004$ the optimal order is
beyond $r=15$. However, for $\epsilon=0.004$ and $\epsilon=0.003$,
the computed remainder values are close to the minimum. The dashed
extensions of the numerical curves shown in Fig.{\ref{fgremopt}}b
correspond to an extrapolation obtained by quadratic fitting of the
available numerical points near the corresponding minima. Using
this extrapolation, we obtain an estimate of the optimal remainder
size for the values $\epsilon=0.004$ and $\epsilon=0.003$, that
will be used in some calculations below. On the other hand,
for $\epsilon=0.002$ and $\epsilon=0.001$, even using the
extrapolation we find that the optimal normalization is beyond
any reliable possibility to estimate given our computing
limitations.

As discussed above, the estimate $r_{opt}\propto 1/\epsilon^{1/2}$
holds \cite{posh1993}, i.e. $r_{opt}$ is expected to be a decreasing
function of $\epsilon$. Fig.\ref{fgremopt}c shows the numerical
estimate for $r_{opt}$ as a function of $\epsilon$ from the points
of minima of Fig.\ref{fgremopt}b. The blue curve is a power-law
fitting, yielding the exponent 0.52, i.e. very close to the one
predicted by theory.

Since the value $||R_{opt}||=||R^{(r_{opt})}(J^{(r)},
\phi^{(r)})||_{\leq 17,{\cal W}_{I_*,\epsilon^{1/2}\rho_0}}$
depends on $\epsilon$, from the above procedure we obtain numerically
pairs of values $(\epsilon, ||R_{opt}||(\epsilon))$. In subsection 3.4
below, we will numerically calculate the value of the diffusion
coefficient $D$ for each one of the selected values of $\epsilon$,
thus allowing for a probe of the dependence of $D$ on $||R_{opt}||$
and a comparison with the results of subsection 2.3.1.

\subsection{Resonant structure}
\begin{figure}
\centering
\includegraphics[scale=0.7]{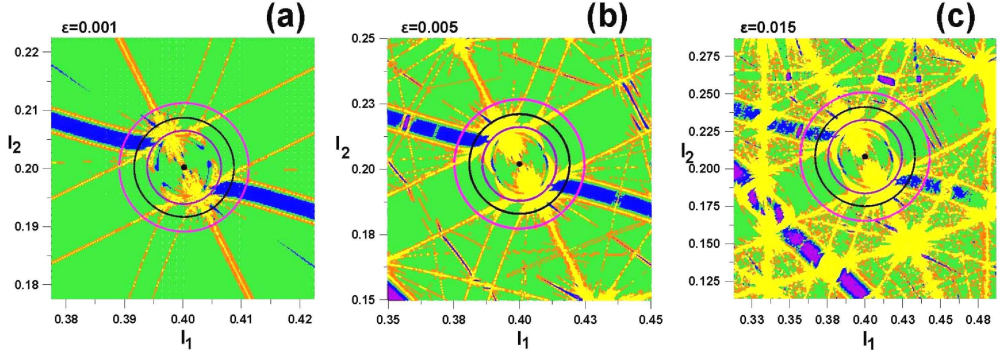}
\caption{FLI map in the action space (surface of section $(I_1,I_2)$
of the Hamiltonian \ref{hamfr} for $\phi_3=0$, $|\phi_1|+|\phi_2|=0$,
around the doubly-resonant point $(I_{1*},I_{2*})=(0.4,0.2)$ for
(a) $\epsilon=0.001$, (b) $\epsilon=0.005$, (c) $\epsilon=0.015$.
The color scale represents the computed value of the FLI (see text)
in the intervals $2\leq FLI< 3$ (magenta, most ordered), $3\leq FLI< 3.5$
(blue), $3.5\leq FLI< 4$ (green), $4\leq FLI<5$ (orange), $5\leq FLI$
(yellow, most chaotic).
}
\label{fgdbleres}
\end{figure}

The resonant structure in the action space (around $I_*$) can
be visualized by employing the method of the FLI map as in
\cite{froetal2000,legetal2003}. We recall that the Fast Lyapunov
indicator (FLI) is a numerical indicator of chaos, defined for
one orbit by
\begin{equation}\label{fli}
FLI = \log_{10}|\xi(t)|
\end{equation}
where $\xi(t)$ is a deviation vector, i.e. in our case
$\xi(t)\equiv(\Delta\phi_1(t), \Delta\phi_2(t),\Delta\phi_3(t),
\Delta I_1(t),\Delta I_2(t),\Delta I_3(t))$ found after solving
the variational equations of motion up to the time $t$ from some
initial conditions $\xi(0)$. By properly choosing a threshold value
$FLI_0\sim\log_{10} t$, orbits with $FLI<FLI_0$ are characterized
as regular, and those with $FLI>FLI_0$ as chaotic. Furthermore, a
convenient use of the FLI in the visualization of the Arnold web
is found by producing FLI color maps \cite{froetal2000}.
Considering a grid of initial conditions in the action space,
we assign to each initial condition a color corresponding
to the FLI value found for the resulting orbit integrated
up to a sufficiently long time (of the order 100 -- 1000
periods). This allows for illustrating the resonant structure
in action space, as shown in Fig.\ref{fgdbleres}, which is
an FLI map in an action domain including our chosen doubly-resonant
point $(I_{1*},I_{2*})=(0.4,0.2)$  for three different values of
$\epsilon$. In all three panels, there are resonances projecting
on $(I_1,I_2)$ as single yellow or orange thick lines, while other
resonances project as strips with a green or blue interior zone
delimited by pairs of nearly parallel yellow or red lines. As explained
in \cite{eft2008}, this difference is only due to the particular choice
of surface of section ($\phi_3=0$, $|\phi_1|+|\phi_2|\leq 0.1$, similar
to \cite{legetal2003}). Namely, the yellow lines marking all resonances
represent the intersection of the thin separatrix-like chaotic layers
formed around each resonance with the chosen surface of section.
This produces a pair of nearly parallel yellow or orange lines for
any resonance (of the form $k\cdot\omega=0$) whose leading Fourier
coefficient $h_k$ of the resonant term $\exp(ik\cdot\phi)$ in the
original Hamiltonian expansion has a negative real part, while it
produces a single yellow or orange thick line if $Re(h_k)$ is
positive. In the latter case, the domain of regular orbits
inside the resonance has no projection on the chosen surface of
section, while in the former case it projects as a strip of green
or blue color.

When $\epsilon=0.001$ (Fig.\ref{fgdbleres}a), we easily distinguish
four main resonances passing through $(I_{1*},I_{2*})=(0.4,0.2)$.
The biggest resonant domain (green, from bottom left to top right)
corresponds to the resonance $\omega_{1}-2\omega_{2}=0$,
whose corresponding wave-vector is the basic resonant wavevector
$k^{(1)}$. Similarly, the single yellow-red thick line going
from bottom right to top left is the resonance
$2\omega_{1}+\omega_{2}-\omega_{3}=0$, whose corresponding
(also basic) wavevector is $k^{(2)}$. We also clearly distinguish
two resonances of order $|k|=5$, namely
$\omega_{1}+3\omega_{2}-\omega_{3}=0$ (blue), and
$3\omega_{1}-\omega_{2}-\omega_{3}=0$ (green). Many other
higher order resonances cross the central doubly-resonant
point $(I_{1*},I_{2*})=(0.4,0.2)$, denoted hereafter by O,
but they are not so visible in the scale of  Fig.\ref{fgdbleres}a.

The resonant strips of all previous resonances join each other
forming a domain of double resonance around O. The extent of this domain
can be determined roughly by drawing concentric circles around the point
O. Such circles correspond to nearly constant normal form energy values,
as can be seen by noting that, for the particular Hamiltonian
function (\ref{hamfr}), the coefficients $a_{ij}$ of Eq.(\ref{hamresn12})
have the values $a_{11}=5$ $a_{22}=5$, and $a_{12}=a_{21}=0$. Applying
Eqs.(\ref{z02},\ref{ene0},\ref{jr0}) for the particular
resonant wavevectors given by (\ref{resvectors}), the
doubly-resonant normal form of the Hamiltonian (\ref{hamfr})
expressed in resonant variables takes the form
\begin{equation}\label{ene02fr}
Z=c_*+6J_F
+{5\epsilon^{1/2}\over 2}\left(J_{R_1}^2+(J_{R_2}+J_F)^2\right)
+O(\epsilon)
\end{equation}
where the $O(\epsilon)$ terms are trigonometric polynomials
of the resonant angles $\phi_{R_1}=\phi_1-2\phi_2$,
$\phi_{R_2}=2\phi_1+\phi_2-\phi_3$, while $c_*=\epsilon^{1/2}0.28186...$
is a constant which appears only in the numerical values of the
quantity
\begin{equation}\label{ez}
E_Z=Z-6J_F
\end{equation}
called, hereafter, the normal form energy ($E_Z$ differs from
the quantity $E'$ defined in Eq.(\ref{eneprime}) only by the
constant $c_*$). We note that the estimate $O(\epsilon)$ for
the trigonometric terms in $Z$ follows from the estimate
(\ref{gn12}) for the size of the corresponding Fourier
coefficients, taking into account that $e^{-\sigma|k^{(1)}|}
\sim e^{-\sigma|k^{(2)}|}\sim\epsilon^{1/2}$, according
to Eq.(\ref{kprime}). Since the angle $\phi_F$ is ignorable in
the hamiltonian (\ref{ene02fr}), $J_F$ is an integral of
the flow of $Z$. Furthermore, since for the particular
choice of Hamiltonian model (\ref{hamfr}) the action $I_3$
is dummy, implying that $I_3$ can be assigned any arbitrary
value without affecting the dynamical evolution of any other
canonical variable, we can always choose the value of $I_3$
so that $J_F=0$. Then, the normal form energy condition
$E_Z=const$ implies
\begin{equation}\label{eres}
{2(E_Z-c_*)\over\epsilon^{1/2}}\equiv\rho^2=5(J_{R_1}^2+J_{R_2}^2)
\end{equation}
where $\rho$ is a $O(1)$ quantity. Transforming to the original
non-scaled action variables we also find
\begin{equation}\label{eres12}
(\epsilon^{1/2}\rho)^2\simeq (I_1-0.4)^2+(I_2-0.2)^2
\end{equation}
whereby $\epsilon^{1/2}\rho$ is interpreted as the radius
of a circle, around O, corresponding to a constant normal form
energy condition. It follows that the set of all possible
normal form energy values are represented on the $(I_1,I_2)$
plane as a set of concentric circles around O. Three such
circles are drawn in Fig.\ref{fgdbleres}a, corresponding to
$\epsilon=0.001$ and $\rho_1=0.31$ (outer circle), $\rho_2=0.27$
(middle circle), and $\rho_3=0.25$ (inner circle). Their main
difference concerns the degree of resonance overlapping in each
case. Namely, for a value of $E_Z=0.0104$, corresponding to the
outer circle $\rho_1=0.31$, the main visible resonances of
Fig.(\ref{fgdbleres}a) intersect the circle in some arcs only,
while the remaining parts of the circle lie in the regular
(non-resonant) domain. In the latter parts, the normal form
dynamics alone would imply the existence of a set of
Kolmogorov-Arnold-Moser invariant tori of large measure.
On the contrary, in the inner circle, corresponding to
$E_Z=0.0099$, all resonances essentially overlap, producing
a strongly chaotic domain. The middle circle corresponds to
$E_Z=0.01007$, which is close to the critical energy below
which resonance overlapping dominates the dynamics.

The remaining panels of Fig.\ref{fgdbleres} show what happens
when $\epsilon$ is increased by a factor 5 ($\epsilon=0.005$,
Fig. \ref{fgdbleres}b), or 15 ($\epsilon=0.015$, Fig.
\ref{fgdbleres}c) with respect to Fig.\ref{fgdbleres}a.
A main feature to notice is that, by increasing
$\epsilon$, many more resonances `show up' in the FLI
map. Furthermore, the size of all resonant domains grows
proportionally to $\epsilon^{1/2}$, as verified in Fig.\ref{fgdbleres},
where by augmenting the scale in panels (b) and (c) by a factor
$\sqrt{5}$ and $\sqrt{15}$ respectively with respect to panel
(a), the widths of all resonant strips passing through $O$ remain
essentially unaltered in all three panels. Thus, the only
essential change is the increase of chaos as $\epsilon$
increases. Namely, we see that the chaotic layers delimiting
the borders of each resonance become thicker as $\epsilon$
increases. This also increases the resonance overlapping
locally, close to the points of resonance crossings.

\begin{figure}
\centering
\includegraphics[scale=0.7]{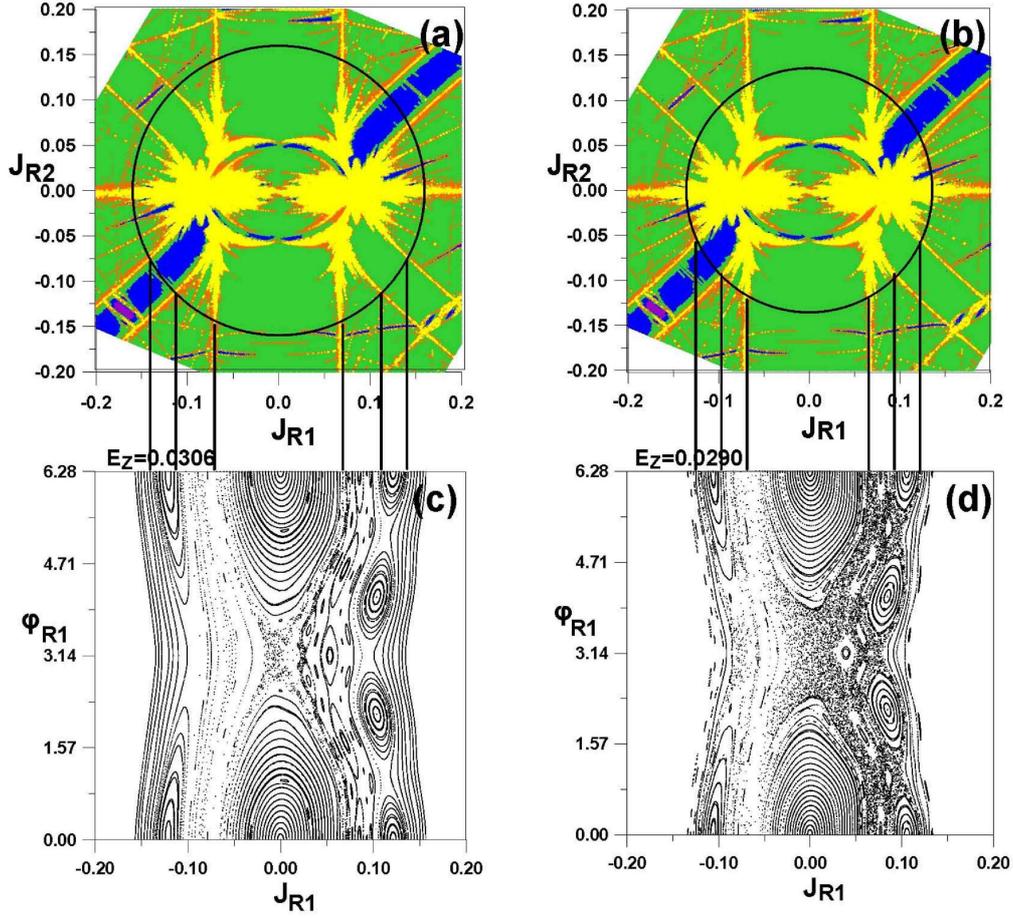}
\caption{(a) and (b) The FLI map in the plane $(J_{R_1},J_{R_2})$
defined as in Eqs.(\ref{restra}) for the Hamiltonian (\ref{hamfr}),
for the same surface of section and using the same color scale as
in Fig.(\ref{fgdbleres}). The circle in (a) corresponds to the
constant normal form energy value $E_Z = 0.0306$ in (a) and
$E_Z=0.029$ in (b). The phase portraits of the normal form
dynamics for the values (c) $E_Z = 0.0306$ and (d) $E_Z = 0.029$.
The plotted surfaces of sections are $(\phi_{R_1},J_{R_1})$ whenever
the quantity $\phi_{R_2}-2\phi_{R_1}=5\phi_2-\phi_3$ (where all
symbols denote the new canonical coordinates and momenta after
the optimal Lie normalization) crosses a multiple value of
$2\pi$. The main resonances are identified as:
$\omega_{1}-2\omega_{2}=0$
(vertical),
$2\omega_{1}+\omega_{2}-\omega_{3}=0$
(horizontal),
$\omega_{1}+3\omega_{2}-\omega_{3}=0$
(bottom left to top right diagonal ),
$3\omega_{1}-\omega_{2}-\omega_{3}=0$
(top left to bottom right diagonal ).
}
\label{fgresportrait}
\end{figure}
Focusing, now, on one value $\epsilon=0.008$, Figure \ref{fgresportrait}
shows in detail the implications of normal form dynamics in the two
regimes when there is no resonance overlap ($E_Z=0.0306$, Figs.
\ref{fgresportrait}a,c), or when there is substantial resonance
overlap ($E_Z=0.029$, Figs.\ref{fgresportrait}b,d). The upper
panels correspond to FLI maps as in Fig.(\ref{fgdbleres}).
Here, however, instead of the action variables ($I_1,I_2$) we
use the resonant re-scaled actions $(J_{R_1},J_{R_2})$, defined
as in Eq.(\ref{restra}), where, for each point in the action
space of the original variables we compute the values of the
transformed actions $J^{(r_{opt})}_i$, $i=1,2,3$ by the composition
of the Lie canonical transformations resulting from the computer-algebraic
program calculating the optimal normal form $Z$. Since the same
program renders also the algebraic form of $Z$, we use this
expression to derive the Hamiltonian equations of motion of
the normal form alone, namely
$\dot{\phi_{R_1}}=\partial Z/\partial J_{R_1}$,
$\dot{\phi_{R_2}}=\partial Z/\partial J_{R_2}$,
$\dot{J_{R_1}}=-\partial Z/\partial \phi_{R_1}$,
$\dot{J_{R_2}}=\partial Z/\partial \phi_{R_2}$,
$\dot{\phi_{F}}=\partial Z/\partial J_{F}$,
while we set $J_F=const=0$. For each value of $E_Z$, we then
compute {\it numerical orbits under the normal form dynamics}
alone via the previous equations. Finally we plot a convenient
surface of section of the normal form flow, taken by the
condition  $mod(\phi_{R_2}-2\phi_{R_1},2\pi)=
mod(5\phi_2-\phi_3,2\pi)=0$. These sections are shown in
Figs.\ref{fgresportrait}c,d, for the normal form energy values
$E_Z=0.0306$ and $E_Z=0.0290$ respectively. The corresponding
circles, through Eq.(\ref{eres}), are shown in panels (a) and
(b), superposed to the color background yielding the FLI map
in the resonant action variables for $\epsilon=0.008$.
The main feature of this plot is the exact correspondence
between the values of $J_{R_1}$ where each resonance intersects
the circle corresponding to $E_Z=const$ in panels (a) and (c),
and the projection of these values to thin chaotic layers
delimiting the same resonance in the corresponding surface of
section. In fact, inside each resonance we have regular orbits
corresponding to islands of stability on the surface of section.
Furthermore, while at the normal form energy value $E_Z=0.0306$
there are many rotational KAM tori separating these resonances,
at the value $E_Z=0.029$ these tori are destroyed and substantial
resonance overlap takes place. This fact leads to the creation
of a connected chaotic domain surrounding all main resonances
in the surface of section of Fig.\ref{fgresportrait}d. This, in turn,
implies that under the normal form dynamics alone no communication
is allowed from one resonance to the other for the normal form
energy value $E_Z=0.0306$ (which in this approximation remains
constant in time), while such communication is possible throughout
the whole connected chaotic domain for $E_Z=0.029$. In fact, the phase
portrait of Fig.\ref{fgresportrait}d renders visually clear that chaos
is rather strong in this case. However, as emphasized in section 2,
this fact has no consequences regarding the possibility of long
excursions in the action space, since all motions in this approximation
would be bounded on circles like those of Figs.\ref{fgresportrait}a,b.
On the contrary, such excursions are only possible due to the effect
of the remainder, which causes the chaotic orbits to slowly `drift'
from circle to circle as the value of $E_Z$ changes slowly in time.
To this we now turn our attention.

\subsection{Visualization of Arnold diffusion in doubly-resonant
normal form variables}
\begin{figure}
\centering
\includegraphics[scale=0.7]{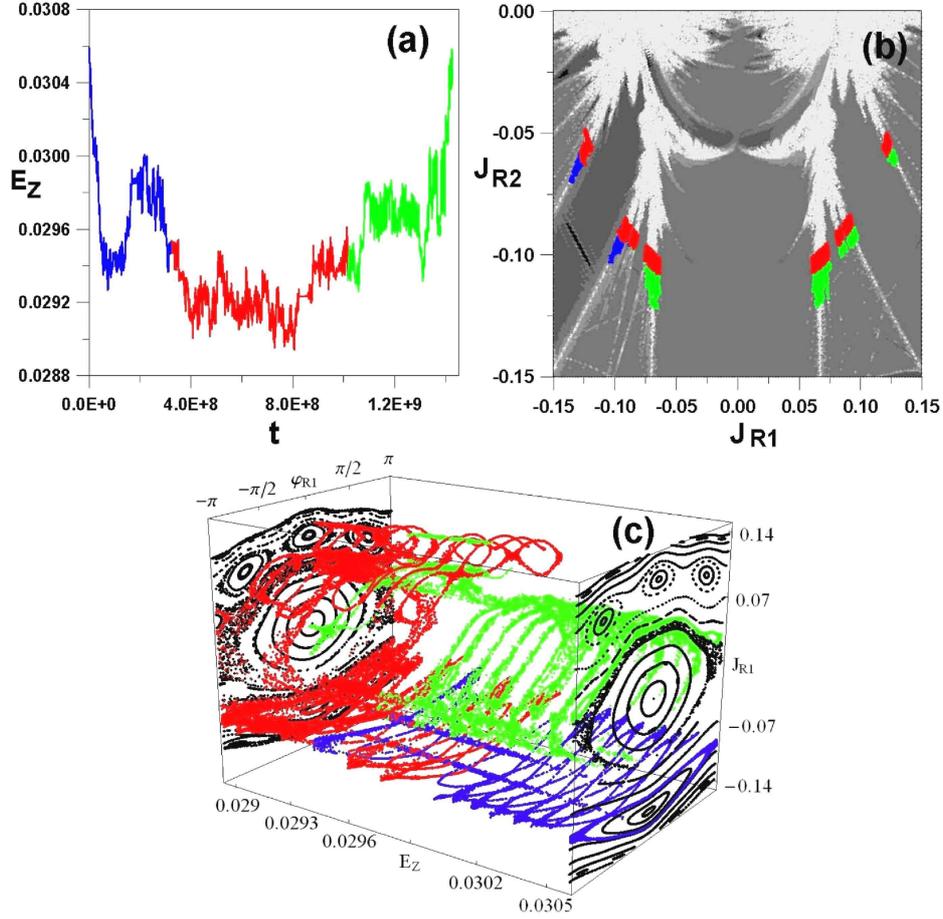}
\caption{Visualization of Arnold diffusion in appropriate variables
of the doubly-resonant normal form, for a numerical orbit in the
Hamiltonian (\ref{hamfr}) for $\epsilon=0.008$. After computing
the optimal normal form, we find, via the Lie canonical
transformations, the values of all transformed variables
$J_{R_1}(t),J_{R_2}(t),J_F(t)$ and $\phi_{R_1}(t),
\phi_{R_2}(t),\phi_F(t)$ corresponding to particular values
of the old variables $J_1(t),J_2(t),J_3(t)$ and
$\phi_1(t),\phi_2(t),\phi_3(t)$ stored at many different times
$t$ within an interval $0\leq t\leq 1.5\times 10^9$ along the
numerical run. Using the numerical values of the computed transformed
variables, (a) shows the variation of the normal form energy $E_Z(t)$
as a function of $t$ in the intervals $0\leq t\leq 3\times 10^8$ (blue),
$3\times 10^8\leq t\leq 10^9$ (red), and
$10^9\leq t\leq 1.5\times 10^9$ (green).
The initial and final values are equal to
$E_Z(t=0)=E_Z(t=1.5\times 10^9)=0.0306$, while the
minimum value, occurring around $t=8\times 10^{8}$ is
$E_Z=0.029$. (b) The evolution of the orbit in the action
space $(J_{R_1},J_{R_2})$, using the same colors as in (a)
for the corresponding time intervals. In the first time interval
(blue), the orbit wanders in the thin chaotic layer of the resonance
$\omega_{1}+3\omega_{2}-\omega_{3}=0$.
In the second time interval (red) it jumps first to the
resonance $3\omega_{1}-\omega_{2}-\omega_{3}=0$,
and then to the resonance $\omega_{1}-2\omega_{2}=0$.
In the third time interval (green) the orbit recedes from the
doubly-resonant domain along the resonance
$\omega_{1}-2\omega_{2}=0$.
(c) 3D plot in the variables $(\phi_{R_1},J_{R_1},E_Z)$,
visualizing Arnold diffusion for the same orbit. Taking
20 equidistant values of $E_{Z,i}$, $i=1,2,\ldots 20$ in the
interval $0.029\leq E_Z\leq 0.0306$, we first find the times
$t_i$ in the interval $0\leq t\leq 9\times 10^8$ when the
normal form energy value $E_Z(t)$ of the numerical orbit
approaches closest to the values $E_{Z,i}$. For each $i$,
starting with the momentary values of all resonant variables
at $t_i$, we then compute 1000 Poincar\'{e} consequents of the
normal form flow on the same section as in
Figs.\ref{fgresportrait}c,d. The same procedure is repeated
in a second interval $9\times 10^8\leq t\leq 1.5\times 10^9$.
As a net result, the orbit at the beginning and end of the
calculation is found on the same section (corresponding to
$E_Z=0.0306$), but in a different resonant layer, having
by-passed the barriers (invariant tori of the normal form
dynamics) via a third dimension (here parameterized by the
time-varying value of $E_Z$).}
\label{fgarnolddif}
\end{figure}

The main effect, of local diffusion within the doubly-resonant domain,
can now be demonstrated with the help of Figure \ref{fgarnolddif}.
The time evolution of one chaotic orbit is shown in this figure,
as the orbit moves within the doubly-resonant domain along some
of the main intersecting resonances. In this example as well we
take $\epsilon=0.008$ (as in Fig.\ref{fgresportrait}).

In Fig.\ref{fgarnolddif}, the evolution of the orbit is shown for
a total time $t=1.5\times 10^9$.
The optimal normal form for $\epsilon=0.008$ has also been computed,
whose optimal normalization order is $r_{opt}=12$, corresponding to
an optimal Fourier order $K_{opt}=36$. Since the corresponding
Lie generating functions are known, we compute, via the composition
of Lie canonical transformations, the values of all transformed
variables $J_{R_1}(t),J_{R_2}(t),J_F(t)$ and $\phi_{R_1}(t),
\phi_{R_2}(t),\phi_F(t)$ corresponding to particular values of
the old variables $J_1(t),J_2(t),J_3(t)$ and $\phi_1(t),
\phi_2(t),\phi_3(t)$ stored at many different times during
the numerical run, i.e. as $t$ varies within the
interval $0\leq t\leq 1.5\times 10^9$. Finally, since the exact
algebraic expression for the normal form $Z$ is known, we compute
the precise numerical value of the normal form energy $E_Z(t)$
at the same times.

Fig.\ref{fgarnolddif}a shows the variation of the normal form energy
$E_Z(t)$ as a function of the time $t$ in the intervals $0\leq t\leq
3\times 10^8$ (blue), $3\times 10^8\leq t\leq 10^9$ (red), and
$10^9\leq t\leq 1.5\times 10^9$ (green). The final time is such that
the initial and final values of $E_Z$ are equal, namely
$E_Z(t=0)=E_Z(t=1.5\times 10^9)=0.0306$. On the other hand, as $E_Z$
slowly changes during the run, it acquires a minimum value around
$t=8\times 10^{8}$, which is $E_{Z,min}=0.029$. Such evolution
corresponds to the process described schematically in
Fig.\ref{fgdbleresmodel} (section 2). Namely, from the previous
figure (Fig.\ref{fgresportrait}) we conclude that the two extreme
values of $E_Z$ acquired during the numerical run are such that
$E_Z(t=0)>E_{Zc}$ while $E_{Z,min}<E_{Zc}$, where $E_{Zc}$ is the
critical energy corresponding to a large scale overlapping of
resonances (subsection 2.3.1). Furthermore, as we will see in the
next subsection, the chaotic excursions of the orbits, and,
consequently, time evolution of $E_Z$, can be approximated by a
normal diffusion process. Furthermore, the fastest evolution takes
place in the intervals $0\leq t\leq 10^8$, and $1.3\times 10^9\leq
t\leq 1.4\times 10^9$, in both of which the total variation of $E_Z$
is of the order of $10^{-3}$, or a `per step' variation of the order
of $\Delta E_Z\sim 10^{-11}$. It should be stressed that these
extremely small variations are possible to unravel numerically only
because we use the new canonical variables deduced by the
normalizing sequence of Lie canonical transformations. When the old
variables are used, instead, we find that the there are large
variations (of order $\epsilon^{1/2}$) of all quantities depending
on the actions. These variations are, in fact, dominated by the
so-called (in the Nekhoroshev theory) `deformation' effects (which
are also of order  $\epsilon^{1/2}$), hence completely covering the
drift effects which are much smaller in size. This feature of the
optimal canonical transformations will be exploited in the
measurement of the diffusion coefficient $D$ as described in the
next subsection.

Fig.\ref{fgarnolddif}b shows the diffusion of the orbit
in the action space $(J_{R_1},J_{R_2})$, using the same colors as
in Fig.\ref{fgarnolddif}a for the corresponding time intervals
(the background produced by the FLI map is shown here in gray
scale). In the first time interval (blue), the orbit wanders
chaotically within the thin chaotic layer of the resonance
$\omega_{1}+3\omega_{2}-\omega_{3}=0$. It should be stressed
that this wandering has a random walk character, i.e. the orbit
makes several reversals of its drift direction, sometimes
approaching and other times receding from the center of the
double-resonance. On average, however, the drift is in the
inward direction (this is a statistical effect; for other
initial conditions the average drift turns to be outwards).
In the second time interval (red), the orbit jumps first to
the domain of the resonance $3\omega_{1}-\omega_{2}-\omega_{3}=0$.
Now, however, the chaotic motion takes place with a relatively
high speed (of order $\epsilon^{1/2}$) in the direction across
resonances. As a result, the orbit fills nearly ergodically
the whole connected chaotic domain surrounding the main
overlapping resonances, while, at the end of this time interval,
the orbit is closer to the resonance $\omega_{1}-2\omega_{2}=0$.
Finally, in the third time interval (green) the orbit recedes
from the doubly-resonant domain (this is also a statistical
effect) being trapped along the domain of the resonance
$\omega_{1}-2\omega_{2}=0$. In this way, at the time
$t=1.5\times 10^9$, the orbit is found at about the same distance
from the center as initially (at $t=0$), but on a different
resonance.

Fig.\ref{fgarnolddif}c, now, shows a 3D plot in the variables
$(\phi_{R_1},J_{R_1},E_Z)$, visualizing the `third dimension'
along which the Arnold diffusion progresses for the same orbit.
From this plot we can clearly see the effect of the remainder,
which can be considered as a very slow modification of the normal
form dynamics acting on a timescale of the order of $10^9$ periods.
The normal form dynamics, on the other hand, describes well the
motion over shorter timescales, of the order of $10^4$--$10^5$
periods. In order to show the dynamical effects happening on both
timescales, we adopt the following numerical procedure: Taking
20 equidistant values of $E_{Z,i}$, $i=1,2,\ldots 20$
in the interval $0.029\leq E_Z\leq 0.0306$, we first find the times
$t_i$ within the interval $0\leq t\leq 9\times 10^8$ (where the
motion is, in general, in the inward direction) when the normal form
energy value $E_Z(t)$ of the numerical orbit approaches the closest
possible to the values $E_{Z,i}$. Then, for each $i$, we set the
momentary values of all canonical variables of the numerical orbit
at the time $t_i$ as initial conditions via which we compute the
corresponding values of all the new resonant canonical variables
following the composition of the corresponding Lie canonical
transformations. With these values as initial conditions, we
compute 1000 Poincar\'{e}  consequents of the normal form flow
alone on the same surface of section as defined in
Figs.\ref{fgresportrait}c,d. The same procedure is repeated in the
second interval $9\times 10^8\leq t\leq 1.5\times 10^9$, where the
motion is in general in the outward direction. The whole set of
Poincar\'{e} consequents (points $(\phi_{R_1},J_{R_1})$ gathered
in this way are plotted in the 2D sections of the parallelepiped
of Fig.\ref{fgarnolddif}c, along with the variations of the value
of the normal form energy $E_Z(t)$ (sampled more frequently) which
are shown in the third dimension.

The details of the filling process of the various resonant chaotic
layers located in the doubly-resonant domain are now clearly seen.
In particular, we note that the chaotic orbit fills the whole
separatrix layer of the initial resonance
$\omega_{1}+3\omega_{2}-\omega_{3}=0$ in a timescale much
shorter than the one required for substantial drift in the
$E_Z$ direction. After a transient `back and forth' motion
around $E_z=0.03$, the orbit then moves slowly towards the
value $E_Z=0.029$, where all important resonances overlap.
In the intermediate time interval (red), we clearly see the
filling of the stochastic layers of both resonances
$3\omega_{1}-\omega_{2}-\omega_{3}=0$ and
$\omega_{1}-2\omega_{2}=0$, while global transport is
allowed by the normal form dynamics from one resonance
to the other. As, however, the remainder effect causes
a new motion of the orbit outwards (i.e. towards higher
values of $E_Z$ (green)), the orbit is eventually captured
at the resonance $\omega_{1}-2\omega_{2}=0$, and stays
there until the end of the simulation at $t=1.5\times 10^9$.

It should be emphasized that the fact that the orbit moves
in the outward direction at $t=1.5\times 10^9$ does not
guarantee that there will be no further return inwards.
In fact, we find that most orbits undergo several `in-out'
cycles like the one described in Fig.\ref{fgarnolddif},
before eventually abandoning the doubly-resonant domain.
As an estimate, for $\epsilon=0.008$ we find that the
number of cycles before a final exit from the doubly-resonant
domain is of the order of 10, while the total time required
for this effect is of the order of $10^{10}$ to $10^{11}$
periods. Furthermore, the probability of exit along one
particular resonance decreases as the order of the resonance
increases. This is expected, since the width of resonances
scales with their order $|k|$ as $\sim e^{-\sigma|k|/2}$,
while the fast filling of the innermost chaotic domains
where all the resonances overlap is nearly ergodic.

 Finally, we point out that a visualization of the diffusion
process like in Fig.\ref{fgarnolddif}c clearly suggests that the
diffusion is driven by the intersections of the asymptotic manifolds
of lower-dimensional objects (like hyperbolic 2D tori) all along the
path in which the diffusion takes place. However, locating  such
tori, and studying their manifolds is a task that cannot be
accomplished by the use of the Birkhoff normal form as above. on the
other hand, the latter provides good initial conditions for a
numerical search of such tori. This subject is proposed for future
study.

\subsection{Dependence of the diffusion coefficient on the optimal
remainder}
Our final goal is to obtain numerical estimates of the value of the
diffusion coefficient $D$ as well as its relation to the size $||R_{opt}||$
of the optimal normal form remainder as $\epsilon$ is varied in the
interval $0.003\leq\epsilon\leq 0.020$.

To this end we implement the following numerical procedure: For any
fixed value of $\epsilon$, using the information from the FLI maps,
we first select 100 initial conditions corresponding to on a circle
defined as in Eq.(\ref{eres12}), where the radius is chosen equal to
$$
\rho=\rho_0=0.27~~.
$$
For such a choice of $\rho$, the corresponding circle lies inside
the resonance overlap domain, ensuring that the short time dynamics
is dominated by the doubly-resonant normal form. However, in longer
times all these orbits exhibit weakly chaotic diffusion.
The complete set of initial conditions for one orbit on the
circle $\rho=\rho_0=0.27$ are found by solving simultaneously
for $I_1$ and $I_2$ the equation of the circle (Eq.(\ref{eres}))
as well as an equation for the initial angle
$\phi_0=arctan[(I_2-0.2)/(I_1-0.4)]$, where, for each initial
condition, $\phi_0$ is chosen by visual inspection so as to
correspond to an initial condition in the domain of each one
of the main overlapping resonances.

\begin{figure}
\centering
\includegraphics[scale=0.7]{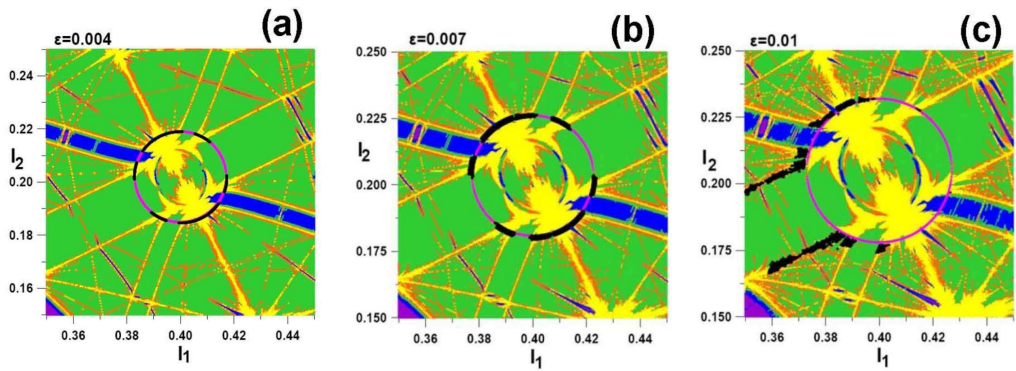}
\caption{Three orbits with initial conditions on the circle
$R(0)=\epsilon^{1/2}\rho_0$ with $\rho_0=0.27$, for
(a) $\epsilon=0.004$, (b)) $\epsilon=0.007$ and
(c) $\epsilon=0.01$. The black points show the orbits' consequents
on the surface of section up to a time $t=10^8$. All three orbits
are diffusing outwards. The circle with radius $R(0)$ is shown in
pink.}
\label{fgdifinout}
\end{figure}
We then follow numerically these orbits for a time long enough
so that the mean change of their radial distance from the center
is large enough to allow for a reliable computation of the diffusion
coefficient. Let $R(t)=\epsilon^{1/2}\rho(t)$ be the instantaneous
value of the distance from the center for any such orbit. The
quantity $[R(t)-R(0)]^2$ changes as an orbit slowly drifts from
one circle to another. Figure \ref{fgdifinout} shows this effect
for three orbits corresponding to the same initial angle $\phi_0$
but for three different values of $\epsilon$, namely
$\epsilon=0.004$ (Fig.\ref{fgdifinout}a),
$\epsilon=0.007$ (Fig.\ref{fgdifinout}b), and $\epsilon=0.01$
(Fig.\ref{fgdifinout}c). The orbits are shown by the black points
on the section $|\phi_1|+|\phi_2|<0.1$, $\phi_3=0$, superposed as
usually to the colored background of the FLI map. The pink circles
in each panel are the circles $R(0)=\epsilon^{1/2}\rho_0$, where
the orbits' initial conditions lie.

Apart from an overall change of the size of the circle of initial
conditions with $\epsilon$, a simple visual comparison of the three
panels suffices to conclude that they imply quite different diffusion
rates of their depicted orbits. In all three panels, the orbits
(black points) are shown up to a time $t=10^8$, which is quite
long compared to the time needed to fill the chaotic domain along
the circle $\rho=\rho_0$. However, when $\epsilon=0.004$
(Fig.\ref{fgdifinout}a), the orbit's plot shows that the orbit
exhibits no discernible transverse motion with respect to this circle,
despite the fact that the orbit lies entirely within a rather strong
chaotic domain (yellow in the FLI scale). On the other hand, when
$\epsilon$ is raised to $\epsilon=0.007$ (Fig.\ref{fgdifinout}b),
the orbit is observed to create a small ring around its initial
circle, implying that the diffusion is visible in this timescale.
Increasing $\epsilon$ still further ($\epsilon=0.01$,
Fig.\ref{fgdifinout}c), causes now a rather fast diffusion,
which leads to the orbit following clearly a preferential
`exit resonance', where the diffusion continues essentially
as in the simple resonance case (subsection 2.3.2).

\begin{figure}
\centering
\includegraphics[scale=0.7]{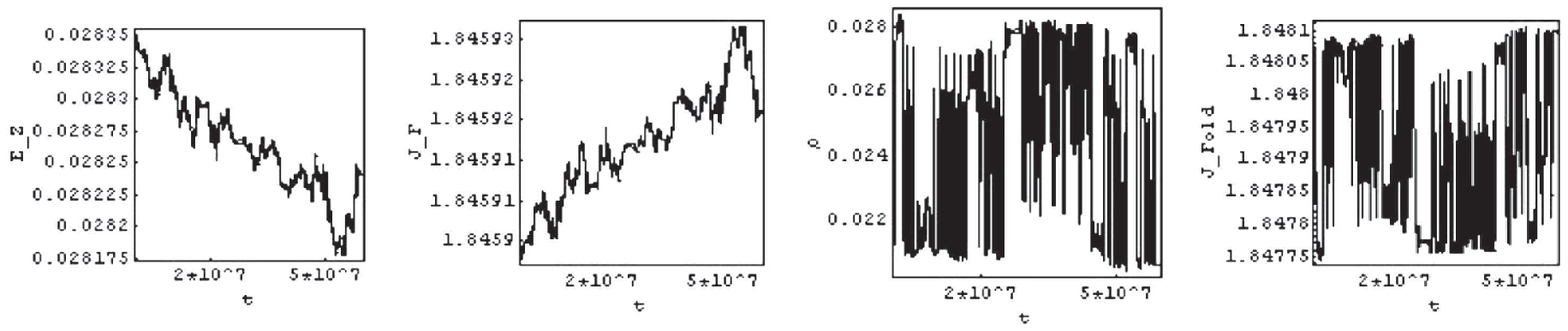}
\includegraphics[scale=0.7]{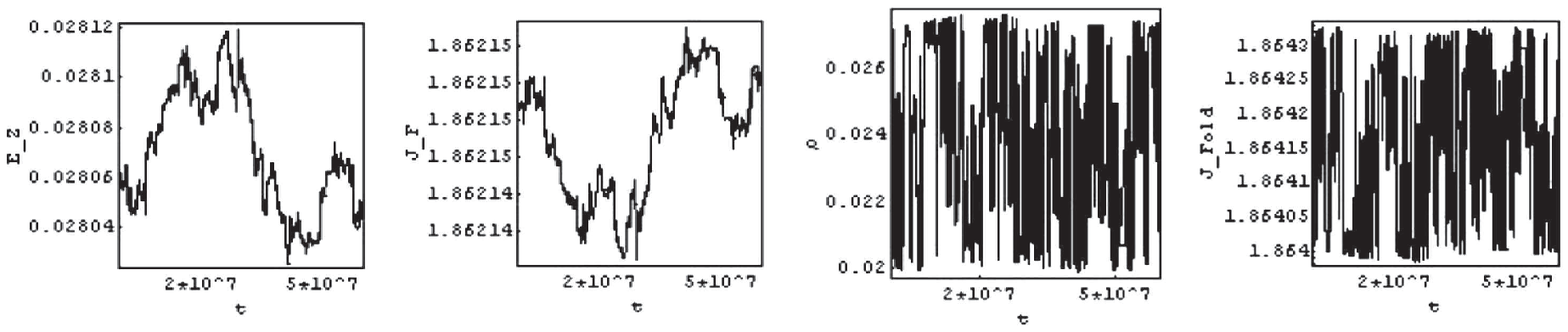}
\caption{The time evolution of the quantities $E_Z$ and
$J_F$ (see text), using the new transformed canonical
variables (first and second panel), and $\rho(t)$ and $J_F$
using the original canonical variables (third and fourth
panel), for two chaotic orbits of our chosen ensemble for
$\epsilon=0.008$ (upper and lower row).}
\label{fgdrift}
\end{figure}
A key remark, now, is the following: similarly to the case of the
orbit of Fig.\ref{fgarnolddif}, whose dynamical features were possible
to unravel using the {\it new, i.e., transformed} canonical variables
after an optimal normalizing transformation, exploiting the same
variables, instead of the original ones, allows to observe
the random walk-like drift of one orbit in the action space
{\it in a much shorter integration time than by the use of the
original variables.} An example is given in Fig.\ref{fgdrift},
for $\epsilon=0.008$. We compute, via the optimal normalizing
canonical transformation, a time sequence of the values of all
the transformed canonical variables $(J^{(r_{opt})}(t),
\phi^{(r_{opt})}(t))$ from the available sequences of values
of the original variables $J(t),\phi(t)$ along the numerical
orbits. The four panels in each row show the time
evolution, for one chaotic orbit on the circle $\rho_0=0.27$,
of the quantities i) $E_Z$ computed in the transformed canonical
variables, ii) $J_F=(2J_1+J_2+5J_3)/30$ computed in the transformed
variables, iii) $\rho(t)$ computed in the original canonical
variables, and iv) $J_F$ computed in the original variables.
We note immediately the gain by passing the data through the optimal
normalizing transformation, namely the fact that this transformation
absorbs all `deformation' effects, allowing to see the very slow
drift due to the weakly chaotic diffusion in a timescale $t\sim
10^7$. In fact, the quantity $E_Z$ can only be computed in the
transformed canonical variables, in which, for both orbits, it
undergoes variations of the order $10^{-4}$. In comparison, the
analog of $E_Z$ in the original variables, i.e., $\rho(t)$,
undergoes variations in the second digit, and the corresponding
time evolution is dominated by $O(\epsilon^{1/2})$ oscillations,
which completely hide the slow drift process in the radial direction
with respect to the central doubly resonant point. The comparison is
even more straightforward in the variables $J_F$ computed by the
transformed and by the original action variables. In the former,
we can clearly see the drift phenomenon for both orbits, which
results in a slow change of the value of $J_F$ (which is an
approximate integral) at the fifth digit. In contrast, this
phenomenon is completely hidden when $J_F$ is computed in the
original variables, since the corresponding plot is dominated by
oscillations of at least one order of magnitude larger amplitude
than the drift effect.

\begin{figure}
\centering
\includegraphics[scale=0.8]{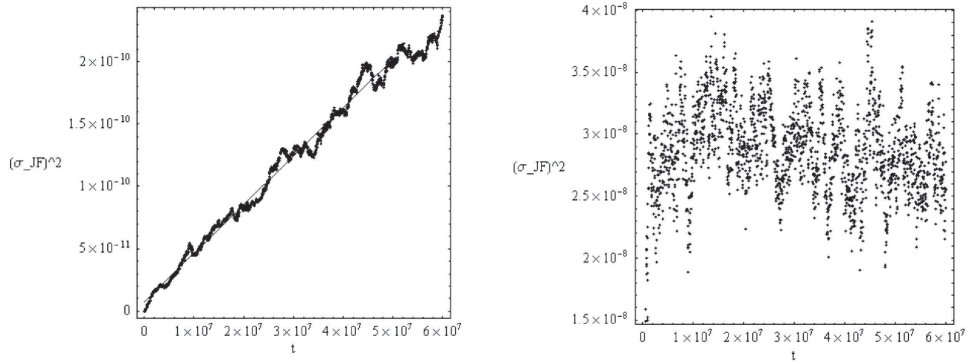}
\caption{The time evolution of the quantity $\sigma_{J_F}^2$ (see
text), in our ensemble of numerical data for $\epsilon=0.008$,
when computed by use of the new transformed canonical variables
(left), or the original canonical variables (right). The evolution
is shown up to the time $6\times 10^7$. }
\label{fgdif008}
\end{figure}
In order, now, to measure the value of the diffusion coefficient,
using the data from all 100 orbits, we define the mean square
deviation:
\begin{equation}\label{dezsq}
\sigma_{y}^2(t)={1\over 100}
\sum_{i=1}^{100} \left(y_{i}(t)-\overline{y}(t)\right)^2
\end{equation}
where $y(t)=Y(t)-Y(0)$, and $Y(t)$ stands for any of the four
quantities shown in Fig.\ref{fgdrift}. Plotting $\sigma_{y}^2$
against the time $t$ allows to estimate the diffusion coefficient.
Figure \ref{fgdif008} shows an example of this calculation, setting
$Y$ equal to $J_F$ in the transformed variables (left panel), or
the original variables (right panel). We note again that it
becomes possible to observe the diffusion in a timescale
$t\sim 10^7$ using the ensemble of data in the transformed
variables, while this time is quite short to reveal any linear
trend of $\sigma_{J_F}^2$ with the time $t$ in the original
variables. In fact, in the original variables it was possible
to measure reliably the diffusion coefficient only after an
integration time $t=10^9$. Furthermore, this time increases
even more for smaller values of $\epsilon$.

\begin{figure}
\centering
\includegraphics[scale=0.5]{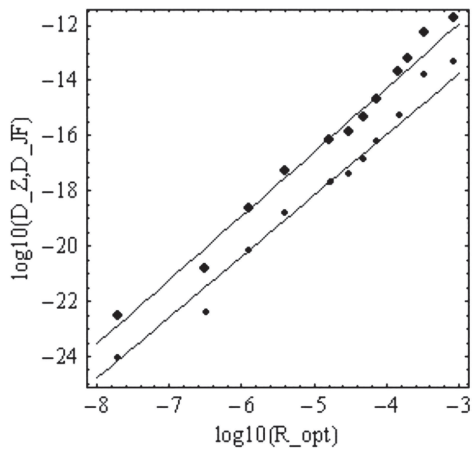}
\caption{Log-log plot of the dependence of the two estimates of the
diffusion coefficient $D_{E_Z}$ (upper set of points) and $D_{J_F}$
(lower set of points) on the optimal normal
form remainder $||R^{(r_{opt})}||$, using numerical data from the
integration of orbits (see text). The points correspond to the values
of $\epsilon$ (from left to right) 0.003, 0.004, 0.005, 0.007,0.008,
0.01, 0.012, 0.013, 0.015, 0.018, 0.020. The straight lines represent
the power-law fits $\log_{10}(D_{E_Z})=-5+2.3\log_{10}(||R_{opt}||)$ (upper)
and $\log_{10}(D_{E_{J_F}})=-7.1+2.2\log10(||R_{opt}||)$ (lower).}
\label{fgdifcfrem}
\end{figure}
Figure \ref{fgdifcfrem} shows the final result. Computing, as
indicated above, the diffusion coefficients $D_{E_Z}$ and $D_{J_F}$
in the transformed canonical variables, for eleven different values
of $\epsilon$ as noted in the caption, we also use the data from
Fig.\ref{fgremopt}, whereby we obtain the optimal remainder value
$||R_{opt}||$ for the same values of $\epsilon$ (from the minima
of the curves of Fig.\ref{fgremopt}). We then plot $D_{E_Z}$ and
$D_{J_F}$ against $||R_{opt}||$ in a log-log scale. Despite some
scatter, the correlation of both independent estimates of the
diffusion coefficient with $||R^{(r_{opt})}||$ can be described
as a power-law. The power-law exponents found by best-fitting are
$p=2.3$ for the data of $D_{E_Z}$ and $p=2.2$ for the data of
$D_{J_F}$. In these best fittings we excluded the two points
for $\epsilon=0.003$ and $\epsilon=0.004$, since the value of
the optimal remainder found by extrapolation is uncertain for
these values of $\epsilon$. However, we note that the corresponding
points in Fig.\ref{fgdifcfrem} are still very close to the fitting
law found by the remaining data.

The exponents found in Fig.\ref{fgremopt} are not far from the
theoretical estimate $p=2$ derived in section 2 (Eq.(\ref{difcfest})).
However, we have made various trials to determine $p$ via alternative
definitions of the diffusion coefficient, and we always find
estimations of $p$ somewhat larger than 2. We thus conjecture
that this difference from $p=2$ is a real effect (not due to
numerical uncertainties), which, however, requires a more
detailed theory to interpret. On the other hand, the corresponding
analysis for simple resonances (subsection 2.3.2) as well as the
numerical results of \cite{eft2008} indicate that the steepening of
the power law in simple resonances of order not smaller than $K'$
is quite substantial, leading closer to $p\simeq 3$. In the latter
case, another independent example \cite{cinetal2013} yields
$p\approx 2.5$. The issue of how exactly to quantify the steepening
of the power-law remains open.

\section{Conclusions}

We examined in detail the phenomenon of weak chaotic diffusion in
doubly or simply resonant domains of Hamiltonian systems of three
degrees of freedom satisfying the necessary conditions for the
holding of the Nekhoroshev theorem. The aim was to determine a
quantitative relation between the diffusion coefficient $D$ and the
size of the optimal remainder $||R_{opt}||$ of a resonant normal
form constructed according to the requirements of the analytical
part of the Nekhoroshev theorem. Our main results are the following:

1) We propose an efficient algorithm for Hamiltonian normalization,
which is implemented as a computer algebraic program performing
expansions up to a high order. We explain the practical aspects
of this algorithm, and show how it can be used in order to compute
i) the optimal normalization order $r_{opt}$ as a function of
the small parameter $\epsilon$, and ii) an estimate of the size
of the remainder $||R_{opt}||$ at the order $r_{opt}$. The
dependence of $r_{opt}$ on $\epsilon$ is found to be an inverse
power-law with an exponent in agreement with theory.

2) We construct estimates on the speed of diffusion in doubly
resonant domains. To this end, we examine first the dynamics
under the Hamiltonian flow induced by the normal form alone
(i.e. neglecting the remainder). The role of the convexity
conditions assumed for the original Hamiltonian is analyzed
in the context of the normal form dynamics. We then discuss
the influence of the remainder on dynamics. Estimates on
the value of the diffusion coefficient $D$ are quantified by
considering a `random walk' model for the slow drift of the
value of the normal form energy due to the remainder. The
final prediction is a power-law estimate $D\sim||R_{opt}||^p$
with $p\simeq 2$ in doubly resonant domains.

3) We perform detailed numerical experiments aiming to test
the above predictions, employing the same Hamiltonian model
as in \cite{froetal2000} as well as the `FLI map' method.
Using the information from the computed normalizing canonical
transformations, we propose a convenient set of variables in
which the Arnold diffusion in the doubly resonant domains is
clearly visualized. Furthermore, using ensembles of chaotic
orbits, we make two independent numerical calculations of the
diffusion coefficient $D$ for various values of $\epsilon$.
The relation between $D$ and $||R_{opt}||$ found by the two
calculations is $D\sim||R_{opt}||^{2.2}$ and
$D\sim ||R_{opt}||^{2.3}$ respectively.

4) Finally, we make some theoretical estimates on the relation
between $D$ and $||R_{opt}||$ in simply resonant domains. In this
case, we combine the basic theory developed in \cite{chi1979}
together with estimates given in \cite{morgio1997} regarding
the dependence of the size of the separatrix splitting on the
optimal normal form remainder in simply resonant domains. We
are thus led to the prediction $||R_{opt}||^{2(1+b)}$, where
$b\simeq 1/2$, or $p=2(1+b)\simeq 3$, holding for all simple
resonances of order higher than $K'$, where $K'$ is defined
in Eq.(\ref{kprime}). The latter result interprets the results
obtained in an earlier study \cite{eft2008} by purely numerical
means. \\
\\
\noindent {\bf Acknowledgements: } We thank two anonymous referees
for a thorough revision of our manuscript, with many constructive
suggestions, as well as Prof. G. Contopoulos for careful reading of
the manuscript. C.E. acknowledges fruitful discussions with the
group of C. Froeschl\'{e}, M. Guzzo and E. Lega.

\appendix

\section{Quasi-convexity and normal form energy constraints}
The quadratic form $\zeta_{0,2}$ given by Eq.(\ref{z02}) can be written as:
\begin{equation}\label{z02matr}
\zeta_{0,2}=(J_{R_1},J_{R_2}) \cdot k^{(1,2)}\cdot M_*
\cdot (k^{(1,2)})^T\cdot (J_{R_1},J_{R_2})^T
\end{equation}
where $k^{(1,2)}$ is a $2\times 3$ matrix whose first and second line
are given by $(k^{(1)}_1,k^{(1)}_2,k^{(1)}_3)$ and
$(k^{(2)}_1,k^{(2)}_2,k^{(2)}_3)$ respectively. Since the matrix $M_*$ is
real symmetric, it can be writen in the form $M_*=X\cdot \mu_*\cdot X^T$,
where $\mu_*=diag(\mu_1,\mu_2,\mu_3)$, with $\mu_i =$ the eigenvalues
of $M_*$, while $X$ is an orthogonal matrix with columns equal to the
normalized eigenvectors of $M_*$. Using the above expression for $M_*$,
Eq.(\ref{z02matr}) resumes the form
$$
\zeta_{0,2} =
(J_{R_1},J_{R_2})\cdot Y\cdot \mu_*\cdot Y^T (J_{R_1},J_{R_2})^T
$$
where $Y=k^{(1,2)}\cdot X$ is a $2\times 3$ matrix. Writing $\zeta_{0,2}$
as $\zeta_{0,2}=Q J_{R_1}^2 + V J_{R_1}J_{R_2} +PJ_{R_2}^2$, and denoting
by $y_{ij}$ the elements of $Y$, the discriminant $\Delta = 4QP-V^2$
is given by:
\begin{equation}\label{z02anal}
\Delta=-[
(y_{11}y_{22}-y_{12}y_{21})^2\mu_1\mu_2+
(y_{11}y_{23}-y_{13}y_{21})^2\mu_1\mu_3+
(y_{12}y_{23}-y_{13}y_{22})^2\mu_2\mu_3]
\end{equation}
Since we have assumed (subsection 2.1) that either all three eigenvalues
$\mu_i$ have the same sign, or two of them have the same sign and one
is zero, by Eq.(\ref{z02anal}) we have that $\Delta<0$. That is, the
quadratic form $\zeta_{0,2}$ is positive definite.


\begin{thebibliography}{}

\bibitem{arn1963}
Arnold, V.I., 1963: {\it Russ. Math. Surveys} {\bf 18}, 9.
\bibitem{arn1964}
Arnold, V.I., 1964: {\it Sov. Math. Dokl.} {\bf 6}, 581.
\bibitem{benetal1985}
Benettin, G., Galgani, L., and Giorgilli, A.: 1985, {\it Cel. Mech.}
{\bf 37}, 1.
\bibitem{bengal1986}
Benettin, G., and Gallavotti, G.: 1986, {\bf J. Stat. Phys.} {\bf 44},
293.
\bibitem{benetal1998}
Benettin G., Fass\`{o} F., Guzzo M.: 1998, {\it Regular Chaot. Dyn.},
{\bf 3}, 56.
\bibitem{ben1999}
Benettin, G.: 1999, in A. Giorgilli (Ed.) `Hamiltonian Dynamics.
Theory and Applications'. {\it Lect. Notes Math.} {\bf 1861}, 1.
\bibitem{cachetal2010}
Cachucho, F., Cincotta, P.M., and Ferraz-Mello, S.: 2010,
{\it Cel. Mech. Dyn. Astron.} {\bf 108}, 35.
\bibitem{celfer1996}
Celletti, A., and Ferrara, L.: 1996,{\it Cel. Mech. Dyn. Astron.}
{\bf 64}, 261.
\bibitem{chi1979}
Chirikov, B.V.: 1979, {\it Phys. Rep.} {\bf 52}, 263.
\bibitem{chivech1985}
Chirikov, B.V., and Vecheslavov, V.V.: 1985, {\it Physica}
{\bf 71}, 243.
\bibitem{cin2002}
Cincotta, P.: 2002, {\it New Astronomy Reviews} {\bf 46}, 13.
\bibitem{cinetal2013}
Cincotta, P., Giordano, C., Mestre, M., and Efthymiopoulos, C: 2013,
in preparation.
\bibitem{con1967}
Contopoulos, G.: 1966, {\it Bull. Astron. CNRS} {\bf 2}, 223.
\bibitem{con2002}
Contopoulos, G.: 2002, {\it Order and Chaos in Dynamical Astronomy},
Springer, Berlin.
\bibitem{dumlas1993}
Dumas, H.S., and Laskar, J.: 1993, {\it Phys. Rev. Lett.} {\bf 70}, 2975.
\bibitem{eftetal1997}
Efthymiopoulos, C., Contopoulos, G. and Voglis, N.: 1998,
in Benest, D., and Froeschl\'{e}, C. (Eds) {\it Discrete Dynamical
Systems}, Gordon and Breach Science Publishers, pp.91-106.
\bibitem{eftetal2004}
Efthymiopoulos, C., Giorgilli, A., and Contopoulos, G: 2004,
{\it J. Phys. A Math. Gen.} {\bf 37}, 10831.
\bibitem{eft2005}
Efthymiopoulos, C.: 2005, {\it Cel. Mech. Dyn. Astron.} {\bf 92}, 29.
\bibitem{eftsan2005}
Efthymiopoulos, C., and Sandor, Z.: 2005, {\it Mon. Not. R. Astron. Soc.}
{\bf 364}, 253.
\bibitem{eft2008}
Efthymiopoulos, C.: 2008, {\it Cel. Mech. Dyn. Astron.} {\bf 102}, 49.
\bibitem{fermel2007}
Ferraz-Mello, S.: 2007, {\it Canonical Perturbation Theories. Degenerate
Systems and Resonance}. Springer (New York).
\bibitem{froetal2000}
Froeschl\'{e}, C., Guzzo, M., and Lega, E.: 2000, {\it Science} {\bf 289}
(5487), 2108.
\bibitem{froetal2005}
Froeschl\'{e}, C., Guzzo, M., and Lega, E.: 2005, {\it Cel. Mech. Dyn. Astron.}
{\bf 92}, 243.
\bibitem{geletal2013}
Gelfreich, V., Sim\'{o}, C., and Vieiro, A.: 2013, {\it Physica D} {\bf 243}, 92.
\bibitem{giocin2004}
Giordano, C.M., and Cincotta, P.M.: 2004, {\it Astron. Astrophys.} {\bf 423}, 745.
\bibitem{gioetal1989}
Giorgilli A., Delshams A., Fontich E., Galgani L., Sim\'{o} C.: 1989,
{\it J. Differ. Eq.}, {\bf 77}, 167.
\bibitem{gioloc1997}
Giorgilli A., and Locatelli, U.: 1997, {\it ZAMP}, {\bf 48}, 220.
\bibitem{giosko1997}
Giorgilli A., and Skokos, C.: 1997, {\it Astron. Astrophys.}, {\bf 317}, 254.
\bibitem{gio1999}
Giorgilli, A.: 1999, In: C. Simo (ed.), {\it Hamiltonian Systems with Three
or More Degrees of Freedom}, Kluwer, Dordrecht.
\bibitem{gio2002}
Giorgilli, A: 2002, {\it Notes on exponential stability of Hamiltonian systems},
in Dynamical Systems. Part I: Hamiltonian Systems and Celestial Mechanics,
Pubblicazioni della Classe di Scienze, Scuola Normale Superiore, Pisa.
\bibitem{gioetal2009}
Giorgilli, A., Locatelli, U., and Sansoterra, M.: 2009, {\it Cel. Mech. Dyn.
Astron.} {\bf 104}, 159.
\bibitem{guzetal2002}
Guzzo, M., Lega, E., and Froeschl\'{e}, C.: 2002, {\it Physica D}
{\bf 163}, 1.
\bibitem{guzetal2005}
Guzzo, M., Lega, E., and Froeschl\'{e}, C.: 2005, {\it Dis. Con. Dyn. Sys. B}
{\bf 5}, 687.
\bibitem{guzetal2006}
Guzzo, M., Lega, E., and Froeschl\'{e}, C.: 2006, {\it Nonlinearity}
{\bf 19}, 1049.
\bibitem{guzetal2011}
Guzzo, M., Lega, E., and Froeschl\'{e}, C.: 2011, {\it Chaos}
{\bf 21}, 033101.
\bibitem{kankon1989}
Kaneko, K., and Konishi, T.: 1989, {\it Phys. Rev. A} {\bf 40}, 6130.
\bibitem{las1993}
Laskar, J.: 1993, {\it Physica} {\bf D67}, 257.
\bibitem{legetal2003}
Lega, E., Guzzo, M., and Froeschl\'{e}, C.: 2003, {\it Physica D}
{\bf 182}, 179.
\bibitem{legetal2007}
Lega, E., Froeschl\'{e}, C., and Guzzo, M.: 2007, {\it Lect. Notes Phys.}
{\bf 729}, 29.
\bibitem{legetal2009}
Lega, E., Guzzo, M., and Froeschl\'{e}, C.: 2009, {\it Cel. Mech. Dyn. Astron.}
{\bf 104}, 191.
\bibitem{legetal2010a}
Lega, E., Guzzo, M., and Froeschl\'{e}, C.: 2010a, {\it Cel. Mech. Dyn. Astron.}
{\bf 107}, 129.
\bibitem{legetal2010b}
Lega, E., Guzzo, M., and Froeschl\'{e}, C.: 2010b, {\it Cel. Mech. Dyn. Astron.}
{\bf 107}, 115.
\bibitem{lhoetal2008}
Lhotka, Ch., Efthymiopoulos, C., and Dvorak, R.: 2008, {\it Mon. Not. R. Astron. Soc.}
{\bf 384}, 1165.
\bibitem{lichlie1992}
Liechtenberg, A.J., and Lieberman, M.A.: 1992, {\it Regular and Chaotic Dynamics},
Springer, Berlin.
\bibitem{loch1992}
Lochak, P.: 1992, {\it Russ. Math. Surv.} {\bf 47}, 57.
\bibitem{math2004}
Mather, J.: 2004, {\it J. Math. Sci.} {\bf 124}, 5275.
\bibitem{mor2002}
Morbidelli, A.: 2002, {\it Modern Celestial Mechanics. Aspects of Solar System
Dynamics}, Taylor and Francis, London.
\bibitem{morguz1997}
Morbidelli, A., and Guzzo, M.: 1997, {\it Cel. Mech. Dyn. Astron.} {\bf 65}, 107.
\bibitem{morgio1997}
Morbidelli, A., and Giorgilli, A.: 1997, {\it Physica D} {\bf 102}, 195.
\bibitem{nei1984}
Neishtadt, A.I.: 1984, {\it J. Appl. Math. Mech.} {\bf 48}, 133.
\bibitem{nek1977}
Nekhoroshev, N.N.: 1977, {\it Russ. Math. Surv.} {\bf 32}(6), 1.
\bibitem{pavguz2008}
Pavlovic, R., and Guzzo, M.: 2008, {\it Mon. Not. R. Astron. Soc.}
{\bf 384}, 1575.
\bibitem{poi1892}
Poincar\'{e}, H: 1892, {\it M\'{e}thodes Nouvelles de la M\'{e}canique
C\'{e}leste}, Gautier-Vilard, Paris.
\bibitem{posh1993}
P\"{o}shel, J.: 1993, {\it Math. Z.} {\bf 213}, 187.
\bibitem{rosetal1966}
Rosenbluth, M., Sagdeev, R., Taylor, J., and Zaslavskii, M.: 1966,
{\it Nucl. Fusion} {\bf 6}, 217.
\bibitem{simval2001}
Sim\'{o}, C., and Valls, C.: 2001, {\bf Nonlinearity} {\bf 14}, 1707.
\bibitem{ten1982}
Tennyson, J.: 1982, {\it Physica D} {\bf 5}, 123.
\bibitem{skoetal1997}
Skokos, C., Contopoulos, G., and Polymilis, C.: 1997, {\it Cel. Mech. Dyn. Astr.}
{\bf 65}, 223.
\bibitem{woodetal1990}
Wood, B.P., Lichtenberg, A.J., and Lieberman, M.A.: 1990,{\it Phys.
Rev. A} {\bf 42}, 5885.

\end{thebibliography}
\end{document}